\documentclass[mancolor,semcolor]{article}
\usepackage{authblk} 
\usepackage{xcolor,soul} 
\usepackage[english]{babel}
\usepackage[letterpaper,top=2cm,bottom=2cm,left=3cm,right=3cm,marginparwidth=1.75cm]{geometry}
\usepackage{amsmath}
\usepackage{graphicx}
\usepackage[colorlinks=true, allcolors=blue]{hyperref}
\usepackage[version=4]{mhchem}

\title{Flat bands, strange metals, and the Kondo effect}

\author[1]{Joseph G.\ Checkelsky}
\author[2,3,4]{B.\ Andrei Bernevig}
\author[5,6]{Piers Coleman}
\author[7]{Qimiao Si}
\author[8]{Silke~Paschen}
\affil[1]{Department of Physics, Massachusetts Institute of Technology, Cambridge, Massachusetts 02139, USA}
\affil[2]{Department of Physics, Princeton University, USA}
\affil[3]{Donostia International Physics Center (DIPC), Paseo Manuel de Lardiz\'abal. 20018, San Sebasti\'an, Spain}
\affil[4]{IKERBASQUE, Basque Foundation for Science, 48013 Bilbao, Spain}
\affil[5]{Center for Materials Theory, Department of Physics and Astronomy, Rutgers University, Piscataway, NJ 08854 USA}
\affil[6]{Department of Physics, Royal Holloway University of
London, Egham, Surrey TW20 0EX, UK}
\affil[7]{Department of Physics and Astronomy, Rice Center for Quantum Materials, Rice University, Houston, TX 77005, USA}
\affil[8]{Institute of Solid State Physics, Vienna University of Technology, 1040 Vienna, Austria}

\begin{document}
\maketitle

\begin{abstract}
\noindent Flat band materials such as the kagome metals or moir\'e superlattice systems are of intense current interest. Flat bands can result from the electron motion on numerous (special) lattices and usually exhibit topological properties. Their reduced bandwidth proportionally enhances the effect of Coulomb interaction, even when the absolute magnitude of the latter is relatively small. Seemingly unrelated to these cases is the large family of strongly correlated electron systems, which includes the heavy fermion compounds, cuprate and pnictide superconductors. In addition to itinerant electrons from large, strongly overlapping orbitals, they frequently contain electrons from more localized orbitals, which are subject to a large Coulomb interaction. The question then arises as to what commonality in the physical properties and microscopic physics, if any, exists between the two broad categories of materials? A rapidly increasing body of strikingly similar phenomena across the different platforms---from electronic localization--delocalization transitions to strange metal behavior and unconventional superconductivity---suggests that similar underlying principles could be at play. Indeed, it has recently been suggested that flat band physics can be understood in terms of Kondo physics. Inversely, the concept of electronic topology from lattice symmetry, which is fundamental in flat band systems, is enriching the field of strongly correlated electron systems where correlation-driven topological phases are increasingly being investigated. Here we elucidate this connection, survey the new opportunities for cross-fertilization in understanding across the platforms, and assess the prospect for new insights that may be gained into both the correlation physics and its intersection with electronic topology.
\end{abstract}

\newpage

\section*{Introduction}
Current studies of quantum materials increasingly focus on the combined role of electron correlations and topology. It has long been recognized that the development of local correlations in electron motion leads to emergent electronic properties such as superconductivity or magnetism. The advent of quantized Hall insulators brought in the realization that quantum matter can also develop a global property: a topology or ``twist'' in the shape and symmetry of the electronic wavefunction which protects new kinds of phases that lie beyond the Landau paradigm of symmetry breaking and obstructs a local representation. Discoveries over the past two decades have uncovered novel phases such as topological insulators and semimetals, and have shown that topology is endemic to band structures and single electron physics \cite{Bra17.1}. The modern interest in correlations and topology is motivated by the expectation that new types of topological phases can develop in correlated systems and that topology will enrich correlation physics. Rapid experimental and theoretical progress across diverse areas of research---from heavy fermion compounds and iron-based superconductors to frustrated lattice compounds exhibiting flat bands and moir\'e superlattices---reveals strikingly similar phenomena. This suggests that universal principles are at play rather than material-specific aspects such as different crystal structures, constituting elements, or degrees of disorder. Here we discuss the prospects this creates for bridging the different areas of research and achieving a more unified understanding.

The appearance of a ``topological flat band'' emerges as one of the common themes across the various materials platforms. It may manifest itself in the kagome flat band system CoSn \cite{Kan20.1},  the candidate fractional Chern insulating state in magic-angle twisted bilayer graphene (MATBG) \cite{Xie21.1,Das21.1} and transition metal dichalcogenides \cite{Cai23.1,Zen23.1,Par23.1}, the heavy Dirac fermions on the surface of the Kondo insulator SmB$_6$ \cite{Pir20.1,Gal19.1}, and the Kondo-driven Weyl nodes \cite{Dzs17.1,Lai18.1} in Ce$_3$Bi$_4$Pd$_3$ \cite{Dzs17.1,Dzs21.1}. Another prominent cross-cutting theme of this new focus of research is strange metal physics. It appears to occur when the correlation strength (or effective band ``flatness'') becomes extreme or even diverges. There is thus a strong basis for breaking the barriers and identifying interconnections across the platforms. In searching for a unifying perspective, we note that in many of these systems there is an interplay of both localized and delocalized electronic states. This suggests that Kondo physics---a prototype means of achieving amplified quantum fluctuations and the strange metal behavior associated with a localization--delocalization transition \cite{Kir20.1,Pas21.1}---may be an overarching principle, in particular when augmented by new topology principles.

\section{Materials platforms}\label{MatPlat}
We start by introducing the materials platforms from which exotic phenomena of strong correlations and topology emerge (FIG.\,\ref{fig1}). Early in the development of quantum mechanics, it was recognized \cite{Wig34.1,Mot37.1} that the properties of the electron fluid are critically dependent on the correlation ratio $U/W$ between their Coulomb ($U$) and kinetic energy (or electronic bandwidth, $W$). At small values of $U/W$, electrons form a weakly interacting Fermi liquid, but at large values, they must localize as an insulator: a Wigner crystal in the electron fluid or a Mott insulator for band electrons. Along the way, new kinds of instability, such as magnetism or even superconductivity might emerge. These early insights still form the basis for the modern research: made practical by new strategies for tuning the correlations, motivated by an ever-growing awareness of the potential for new phases of quantum matter, phases in which both correlations and topology may play a major role.
 
The canonical pathway to strong electron correlations is via materials with highly localized electron orbitals, such as $d$ or $f$ band systems. A more recent strategy for increasing correlations, established in the class of ``flat band'' materials, is quenching the kinetic energy, either via destructive interference on special ``frustrated lattices'' or by artificially enhancing the unit cell as in the case of moir\'e systems. Such flat bands often turn out to be topological. In turn, the long-studied strongly correlated electron systems are enriched by band topology in the form of novel gapless and gapped topological phases. Most prominently in the heavy fermion compounds, extremely narrow (or ``flat'') bands are also at play, raising the prospect of a unified understanding across these seemingly disparate materials classes. 

Recent experimental and theoretical work nourishes this hope. Firstly, there is evidence for heavy fermion behavior in flat band systems, suggesting a common role of charge localization. Secondly, the quintessential strong correlation behavior of strange metallicity is observed across the different platforms, with a likely connection to a (dynamical) charge localization--delocalization transition. Finally, the paradigm of band topology presents obstructions to a real-space representation with exponentially localized orbitals. The interplay between $U$, $W$, and the band topology thus also represents a new ``frustration'' frontier where delocalization competes with localization via correlations and topology.

\subsection*{Flat band systems}
\begin{figure}[ht!]
\centering
\includegraphics*[width=0.9\textwidth]{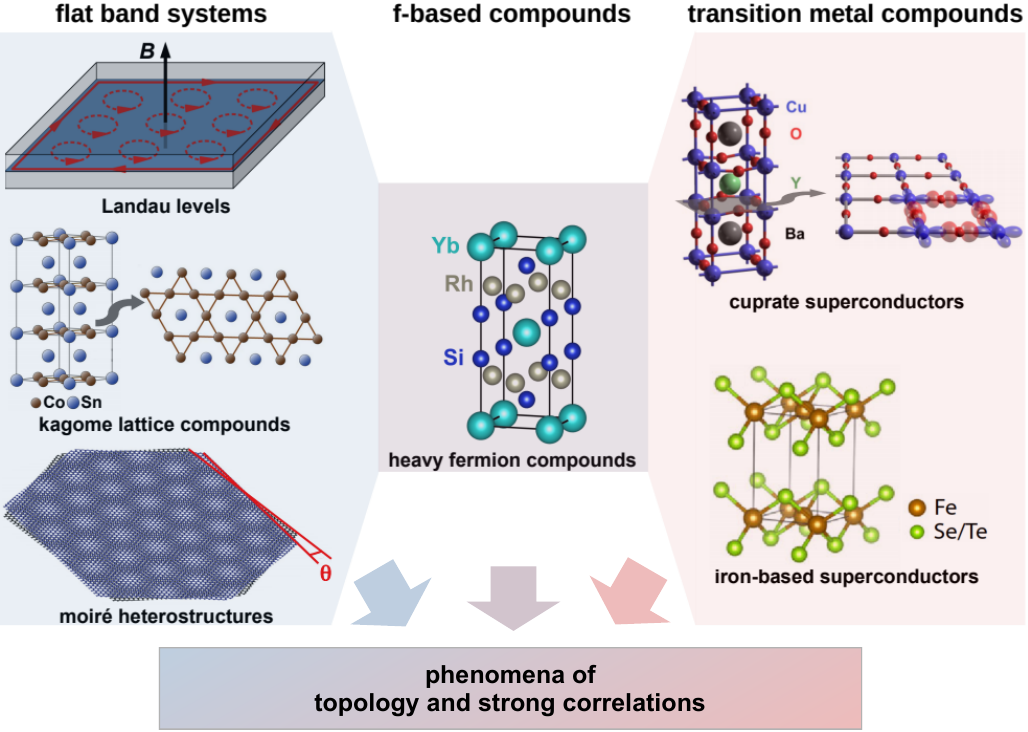}
\caption{\label{fig1} {\bf Platforms for topology and correlation phenomena.} {\bf(left)} Flat band systems: integer quantum Hall (IQH) state in a 2-dimensional electron gas (2DEG) as example of a continuum system (top), the frustrated lattice kagome compound CoSn (middle), and the moir\'e superlattice system MATBG (bottom); {\bf(center)} $f$-based compounds: the heavy fermion compound YbRh$_2$Si$_2$; and {\bf(right)} transition metal compounds: the cuprate YBa$_2$Cu$_3$O$_7$ (top) \cite{Bar13.2} and the Fe-based superconductor Fe(Se,Te) (bottom).}
\end{figure}

In the strictest sense, a flat band has zero energy dispersion and bandwidth. However, as we discuss below, a looser use of the term permits us to develop a broader materials perspective. Landau levels (FIG.\,\ref{fig1} left, top), resulting from the application of a magnetic field perpendicular to a two-dimensional (2D) high-mobility electron gas \cite{Kli80.1,Kli86.1}, are a classic example of flat bands. The quantum Hall effect (QHE) underlying this seminal discovery has since been realized in many different settings \cite{Kli20.1}: thus the observation of a half-integer QHE in graphene \cite{Zha05.1} contributed to the detection of Berry phases and band topology in crystalline matter. A fractional QHE \cite{Sto99.1} also emerges as correlations due to the weak screening in two dimensions become important at very low temperatures. Currently, fractionalized topological states driven by band topology are being pursued, either in small field \cite{Xie21.1,Cai23.1} or more remarkably, in zero field \cite{Zen23.1,Par23.1} as expected for a fractional Chern insulator.

Flat bands may also arise in special lattices (Lieb, kagome, pyrochlore, etc.), which we will refer to as ``frustrated lattices''. This can be achieved in bipartite lattices \cite{Lie89.1,Mie91.1,Sut86.1,Oga21.1,Cal22.1,Nev23.1} with a different number of atoms in the two sublattices and hoppings satisfying special conditions. Line and split graph, as well as chiral flat bands are a subdivision of this general technique. Their flatness, perfect under special conditions (which can be general enough to include many orbitals, spin--orbit coupling, and any distance hoppings), results from destructive interference between different electron paths and thus from extended as opposed to atomically localized states. This flat band concept appears to generalize to 3D materials, though deviations from the special conditions introduce finite dispersion. This has been reported to manifest in the electronic structure of the kagome metal CoSn (FIG.\,\ref{fig1} left, middle), where bands with kinetic energy quenched by nearly an order of magnitude relative to conventional bands therein are observed \cite{Kan20.1}. Systematic searches have in the meantime identified thousands of other candidate materials \cite{Reg22.1,Chi22.1,Cal22.1,Nev23.1}. In the presence of spin--orbit coupling, the flat bands may inherit a $Z_2$ topological character \cite{Reg22.1}.

A third example of flat band realization is that of moir\'e materials \cite{And21.1}, where a superlattice with ``effective'' moir\'e sites and a unit cell much larger than that of the original lattice forms. A large unit cell leads to the appearance of many bands in a finite bandwidth system, thereby flattening the dispersion of isolated bands. This flattening is due to band folding and subsequent level repulsion. Such materials can be realized in van der Waals heterostructures \cite{Gei13.1}, for instance by twisting layers with respect to each other. A prominent example is bilayer graphene when twisted to a so-called ``magic'' angle, which was theoretically predicted \cite{Sua10.1,Bis11.1} and experimentally realised \cite{Cao16.1,Cao18.1,Cao18.2} (FIG.\,\ref{fig1} left, bottom). Also other bi- and few-layer van der Waals materials have been twisted to form moir\'e heterostructures, e.g. bilayers of the transition metal dichalcogenide MoTe$_2$ \cite{Cai23.1,Zen23.1,Par23.1}. In addition, other ways of introducing superlattice potentials are being explored, including the use of lattice mismatch \cite{Reg20.1} and patterned dielectrics to electrostatically define a superlattice potential \cite{For18.1,Gho23.1}.

Finally, flat bands may also be realized in non-electronic settings, e.g.\ in phononic, magnonic, metamaterial, or cold atom systems \cite{Ley18.1}. For instance, a bound state in Floquet dynamics consisting of up to five microwave photons has been realized in a ring of 24 superconducting qubits \cite{Mor22.1}. The bound photons, which behave as quasiparticles with well-defined momentum, energy and charge, were shown to develop a very low velocity as the interaction strength increases with the number of photons. Another example are electric circuit networks, which were shown to emulate tight-binding models of line graph lattices. For instance, a honeycomb lattice was shown to exhibit a Dirac crossing in the bulk admittance band structure and flat band surface states \cite{Hel19.1}.

\subsection*{$f$-based compounds}
The rare earth elements---from Ce to Yb---have partially filled $4f$ orbitals. These orbitals are highly compact and shielded by much larger $5d$ and $6s$ orbitals, so they retain their identity even when part of a tightly packed solid. The wave functions of the fourteen $4f$ states, the degeneracy of which is partially lifted by spin--orbit coupling and crystal electric field effects, have negligible overlap and thus correspond to localized states. In isolation, these can be seen as forming perfectly flat bands---atomic flat bands. If the ground state level, typically a Kramers doublet (but also non-Kramers doublets \cite{Oni16.1}, spin-orbital-entwined quartets \cite{Mar19.1,Liu23.1}, etc.\ are possible), is positioned well below the Fermi energy, the large value of $U$ restricts it to single occupancy, leading to magnetic, unpaired electrons known as ``local moments''. A priori, these local moments can lead to magnetic order, usually mediated by the Rudermann--Kittel--Kasuya--Yosida (RKKY) interaction. However, if the electronic density of states is sufficiently large at the Fermi energy, the conduction electrons magnetically screen these local moments, forming entangled singlets, a process known as the ``Kondo effect'' \cite{Hew97.1,Col15.1,Kir20.1,Pas21.1}. The low-energy excitations of the resulting ground state form a very narrow, though not perfectly flat, heavy fermion band. Also this band can undergo an ordering transition, typically in the form of a spin-density wave.

The class of materials exhibiting such physics are called heavy fermion compounds, with YbRh$_2$Si$_2$ being a prototype example (FIG.\,\ref{fig1} center) \cite{Tro00.1,Ngu22.1}. The term ``heavy'' refers to the strongly enhanced effective masses of the quasiparticles (see Sect.\,\ref{Routes}), manifesting in the low-temperature thermodynamic and transport properties \cite{Ste84.1,Kad86.1,Jac09.1}. Heavy fermions can also be realized in compounds containing actinides, with U-based materials as prominent examples. $5f$ orbitals are less localized than $4f$ orbitals, which typically leads to more modest bandwidth renormalizations. In addition, the involvement of multiple $5f$ orbitals in the low-energy physics leads to orbital-selective correlations, by analogy with what happens in multi-orbital $3d$ electron systems to be discussed below.

It is standard to consider heavy fermions in terms of quasiparticles described by the (non-relativistic) Schr\"odinger equation, with an underlying quadratic ($k^2$) dispersion and a large effective mass that is inversely proportional to the band curvature, giving rise to very narrow bands. Only recently has it been realized that there are heavy fermion compounds that require a description in terms of the (relativistic) Dirac equation. They may result from filling constraints and strong spin--orbit coupling in the form of flat topological surface states in Kondo insulators \cite{Dze16.1,Pir20.1} or, with the cooperation of symmetry, as conductive topological bulk states \cite{Dzs17.1,Lai18.1,Dzs21.1,Che22.1}. The ultralow velocity of the ``Dirac-like'' quasiparticles is then a measure of the width of the ``topological flat band''.

\subsection*{Transition metal compounds}
The spatial extent of atomic orbitals increases further in the transition metal elements. Depending on their arrangement in the solid, the $d$ orbital overlap with other atoms can still be small enough to produce strongly interacting narrow bands, as in the case of $3d$ transition metal oxides.

A prominent example is the cuprate superconductors \cite{Bed86.1} which consist of conducting copper--oxygen planes sandwiched between insulating dopant layers. A square lattice of $d_{x^2-y^2}$ orbitals within these planes (FIG.\,\ref{fig1} right, top) dominates the low-energy electronic states, though the importance of other orbitals (Cu\,$d_{3r^2-z^2}$ and O\,$2p_x$, $2p_y$ \cite{Zha88.1}) has also been considered \cite{Sob21.1}. When undoped, every site contains one hole and the band is half-filled: the large onsite Coulomb repulsion then gives rise to a Mott insulator, rather than the metal expected in a non-interacting description---a strong correlation effect. Doping promotes intersite hopping and, at large doping, the system becomes an essentially normal metal. At intermediate doping, strange metal behavior and unconventional superconductivity develop, likely accompanied by a charge localization--delocalization transition, phenomena of great interest. The discovery of infinite-layer nickelates \cite{Li19.3} has opened up a new playground for this physics. These  systems are generally considered to be topologically trivial, unless perhaps in the form of twisted devices \cite{Can21.2,Vol23.1}.

Another much-investigated family of transition metal compounds is the iron-based superconductors \cite{Joh11.1,Si23.1}, with representative examples given by the quaternary oxypnictide LaOFeAs \cite{Kam08.1} and the tetragonal chalcogenide superconductor $\alpha$-FeSe \cite{Hsu08.1}. Across these families, an Fe-based planar sublattice (FIG.\,\ref{fig1} right, bottom) with two iron atoms per unit cell, is the key building block. Unlike in the cuprates, several orbitals are essential for the low-energy electronic states \cite{Yu21.1}. As discussed below, strange metal \cite{Kas10.1} and topological phenomena \cite{Zha18.2} are observed in some of these materials. Finally, the unifying concept of localization--delocalization and strange metallicity also applies to other $3d$ electron systems such as the heavy-fermion-like compounds LiV$_2$O$_4$ \cite{Ura00.1} and Na$_{1.5}$Co$_2$O$_4$ \cite{Miy04.1}, the $4d$ transition metal oxide Sr$_3$Ru$_2$O$_7$ \cite{Bru13.1}, the organic conductor $\kappa$-(BEDT-TTF)$_4$Hg$_{2.89}$Br$_8$ \cite{Wak23.2}, the 1$T$-phase transition metal dichalcogenides \cite{Che20.1,Van21.1,Liu21.3}, and possibly even to more traditional transition metal magnets \cite{Mor79.1,Von93.1,Hau17.1}.

\section{Different routes to strong correlations}\label{Routes}
In this section we discuss the different pathways to strong correlations---as defined by the ratio $U/W$---across the aforementioned materials platforms. We also describe various experimental methods to quantify the correlation strength.

\subsection*{Coulomb vs kinetic energy effects}
The minimum model for heavy fermion compounds is the periodic Anderson model, containing a lattice of almost-localized $f$ orbitals coupled to $spd$ orbital-based conduction bands with itinerant $c$ electrons. The $f$ orbitals sit at an energy $-\epsilon_f$ relative to the Fermi energy and are subject to an onsite Coulomb repulsion $U$. A hybridization $V$ describes the tunneling of the $f$ electrons into the conduction sea. The repulsive interaction energy between conduction electrons is small compared to their bandwidth $W$ and is typically ignored. In the Kondo regime where $\epsilon_f$ and $U$ are large, the $f$ states become singly occupied, forming magnetic or ``local'' moments. This permits the periodic Anderson model to be mapped to a Kondo lattice model, where the correlations are encoded in an exchange interaction between the local moments of the $f$ states and conduction electron spins. At low temperatures, a Kondo resonance with a width set by the Kondo temperature $T_{\rm K}$ builds up at the Fermi energy as the local moments and conduction electrons form bound-state singlets \cite{Hew97.1}. A large $U/W$ corresponds to a small $T_{\rm K}$.

In many transition metal compounds, electronic states near the Fermi energy are primarily associated with $d$ orbitals and, especially in the most localized $3d$ electron materials, electron correlations are dominated by the onsite Coulomb (or ``Hubbard'') interaction ($U$) and further Coulomb terms when multiple $d$ orbitals are involved \cite{Ima98.1}. In effective one-band models such as those used to describe high-$T_{\rm c}$ cuprates, when $U$ is large compared to the bandwidth $W$ near half-filling, the system becomes a Mott insulator \cite{Ima98.1}: this state anchors the strong correlation physics of the nearby metals and superconductors, which are thus doped Mott insulators. Unlike heavy fermion compounds, the largest values of $U/W$ are thus associated with insulators. For multi-orbital systems such as the iron pnictides or chalcogenides, the Mott transition can become orbitally-selective Mott, allowing metallic behavior to survive at the largest values of $U/W$.

Flat band (topological) behavior can intellectually be traced back to early studies of the quantum Hall effect: a sufficiently large magnetic field applied perpendicular to a 2D electron gas creates Landau levels and quenches the kinetic energy and bandwidth $W$, considerably amplifying $U/W$ so that correlation-driven insulating topological states---the fractional QHE---are realized. Today, flat bands can be engineered in a large number of frustrated lattices and moir\'{e} systems. An extreme example is MATBG, where the moir\'{e} superlattice gives rise to bands so narrow \cite{Bis11.1} that their bandwidth can be lower than the small $U$ of the carbon $p$ electrons. Meanwhile, materials on frustrated lattices allow for the possibility of destructive interference in the electron motion. The associated band will have very small kinetic energy. This route is being explored primarily in transition metal compounds where, in combination with the sizeable $U$ of the $d$ orbitals, large values of $U/W$ can result.

Thus, even though the microscopic descriptions of the strong correlations seem to imply distinct physics across the materials platforms, the barriers are only apparent. The correlation degree is ultimately determined by a dimensionless ratio. A large Coulomb parameter $U$ in the $f$ and $3d$-electron-based compounds or a small bandwidth $W$ in flat band systems has the same physical consequence: $U/W$ is large and thus the correlations are strong.

\subsection*{Experimental measures of correlation strength}
A diverse set of experimental tools has been used to characterize the dimensionless correlation strength $U/W$. While there is a long tradition of benchmarking correlations in topologically trivial materials, similar efforts for systems with topological bands are less established. For Dirac-like quasiparticles, using the Dirac velocity appears as a valid approach.

If, at sufficiently low temperatures, a material behaves as a Fermi liquid, Fermi liquid parameters can be used to extract the mass renormalization $m/m_0$, where $m$ and $m_0$ are the renormalized and free electron mass respectively. Most common are the determination of the Sommerfeld coefficient $\gamma=C_{\rm el}/T$ of the electronic specific heat $C_{\rm el}$ and the $A=(\rho -\rho_0)/T^2$ coefficient of the electrical resistivity $\rho$, where $\rho_0$ is the residual resistivity at $T=0$. In heavy fermion compounds, the renormalizations are so strong that these electronic terms dominate contributions from other excitations such as phonons. This has been explicitly demonstrated by measurements on ``non-$f$ reference materials'', e.g.\ compounds where La replaces Ce. Furthermore, measurements down to a few tens of milli-Kelvin are standard in this field and records to below one milli-Kelvin \cite{Ngu22.1} have been demonstrated. Heavy fermion compounds neatly follow the Kadowaki--Woods scaling $A/\gamma^2\approx 10 \mu\Omega{\rm cm\,mol}^2{\rm K}^2{\rm J}^{-2}$ \cite{Kad86.1,Jac09.1}, which confirms that $\gamma$ and $A$ measure the same quantity, $m/m_0$. Whereas $m/m_0$ can be directly determined from $\gamma$ if the charge carrier concentration $n$ is known (green symbols in FIG.\,\ref{fig2}), the determination of $m/m_0$ from $A$ assumes the above $A/\gamma^2$ ratio (blue symbols). $\gamma$ has also been used to estimate the correlation strength in transition metal oxides and in the frustrated lattice compound Ni$_3$In (further green symbols). Fermi liquids also show a Drude response in the optical conductivity, though the Drude peak may be shifted to extremely low frequencies in strongly correlated systems \cite{Li23.1}. When it is resolved, $m/m_0$ can be extracted from the Drude weight, which is proportional to the so-called optical kinetic energy $K$ \cite{Qaz09.1,Si09.1,Deg11.1} [pink symbols for Ba(Fe$_{1-x}$Co$_x$)$_2$As$_2$].

Also other types of spectroscopy are powerful tools, provided the resolution is sufficient to resolve the renormalized bands and quasiparticle lifetime broadening effects do not smear the bands too strongly. For instance, bands observed in angle-resolved photoemission spectroscopy (ARPES) can be fitted with a Schr\"odinger-like dispersion $\epsilon = \hbar^2 k^2/(2 m)$ to extract $m/m_0$ (orange symbols for FeSe$_{1-x}$Te$_x$). This approach can be extended to materials with Dirac-like dispersion, $\epsilon = \hbar v k$, where the measured Dirac velocity $v$ can be scaled to the velocity $v_0$ of a generic non-interacting system, e.g.\ $10^6\,{\rm m/s}$ of graphene (violet symbols for $T_d$-TaIrTe$_4$ and $T_d$-MoTe$_2$). For systems with stronger correlations, quasiparticle interference patterns in scanning tunneling microscopy (STM) can be used, in particular to detect surface states (violet symbols for SmB$_6$). Band-specific information has also been extracted from quantum oscillation experiments, by fitting the temperature dependence of the amplitude of Shubnikov--de Haas (SdH) oscillations of the electrical resistivity (violet symbols for twisted bilayer graphene) or de Haas--van Alphen (dHvA) oscillations of the magnetization [orange symbols for BaFe$_2$(As$_{1-x}$P$_x$)$_2$] to the Lifshitz--Kosevich formula \cite{Sch84.1}; as in the case of specific heat, this requires knowledge of the carrier density $n$. These band-resolved techniques measure renormalizations of single bands, which may or may not represent overall renormalizations (such as probed by $\gamma$ and $A$). In the case of extreme renormalization, $(v/v_0)$ can be directly extracted from specific heat (violet point for Ce$_3$Bi$_4$Pd$_3$).

In many correlated materials, non-thermal control parameters such as pressure, magnetic field, or doping have been shown to tune the correlation strength. Interestingly, when the correlation strength reaches extreme values for a given system, a totally different type of behavior may emerge: ``strange metallicity'' (see star symbols in FIG.\,\ref{fig2}), a particular type of non-Fermi liquid behavior, which we discuss next.

\section{Strange metal behavior}\label{StrMet}
A vast body of experiments on the cuprate superconductors reveals strange metal behavior at or near ``optimal doping'' in the generic temperature--doping phase diagram of the cuprates \cite{Kei15.2}, i.e., where the transition temperature for superconductivity reaches its maximum. This concurrance of strange metallicity and optimum superconductivity is also seen in heavy fermion compounds when tuned by (external or chemical) pressure \cite{Tau22.1}. In contrast to the cuprates, which undergo a Mott metal--insulator transition at small doping, away from the strange metal regime, strange metal heavy fermion compounds remain metallic across the entire accessed range of experimental tuning parameters, including doping, pressure, and magnetic field. Nevertheless, Mott physics is considered to play a key role in generating strange metal physics, albeit in the form of an ``orbital-selective'' Mott transition, manifested as a ``Kondo destruction transition'' \cite{Kir20.1,Pas21.1}. Here we survey the broad range of materials outlined in FIG.\,\ref{fig1}, using the notion of Kondo destruction and its suitable generalizations to provide an anchoring perspective on strange metallicity.

\subsection*{Kondo destruction and electronic localization}
In heavy fermion metals, the Kondo interaction promotes the formation of a Kondo singlet, which is responsible for the low-energy delocalization of the $f$ electrons. At the same time, the antiferromagnetic RKKY interaction promotes the development of spin singlets among the local moments; this not only leads to a tendency towards magnetic order but also competes with the Kondo singlet formation, ultimately giving rise to a destruction of the Kondo singlets and a localization of the $f$ electrons, in a dynamical process associated with strong quantum fluctuations. It is quite remarkable that related properties have also been observed in other materials across the platforms of FIG.\,\ref{fig1}.

\subsection*{Linear-in-temperature electrical resistivity}
Strange metallicity is most frequently associated with an unusual linear-in-temperature electrical resistivity (FIG.\,\ref{fig3}a) that appears instead of the expected Fermi liquid $T^2$ form and was first highlighted decades ago in the cuprate high-temperature superconductors \cite{Cav87.1,Cav87.2}. Examples of materials revealing this behavior are found across all three platforms of FIG.\,\ref{fig1}, as highlighted by the cuprate La$_{2-x}$Sr$_x$CuO$_4$ \cite{Coo09.1} as member of the transition metal compounds (FIG.\,\ref{fig3}b); the heavy fermion metal YbRh$_2$Si$_2$ representing the $f$-based compounds (FIG.\,\ref{fig3}c), where the temperature exponent 1 spans more than 3 orders of magnitude in temperature \cite{Ngu21.1}; flat band systems such as the kagome metal Ni$_3$In \cite{Ye21.1x,Ye23x} (FIG.\,\ref{fig3}d) and MATBG \cite{Jao22.1} (FIG.\,\ref{fig3}e). The appearance of linear resistivity across all three platforms is remarkable and suggests a common underlying principle. Theories based on distinct starting points---from Kondo destruction models \cite{Si01.1,Col01.1,Sen04.1} to SYK-motivated models \cite{Pat19.1,Guo20.1,Chr22.1} and ersatz Fermi liquids \cite{Els21.2}---have been proposed to explain or be compatible with this behavior. However, it is difficult to pin down the correct description based on electrical resistivity data alone and thus all salient features associated with it should be considered. Finally, to explicate on the role of disorder in the strange metal physics, we note that this physics develops in stoichiometric materials---including heavy fermion metals---that are very clean. For example, a residual resistance ratio of 150 and a residual resistivity of $0.5\,\mu\Omega$cm have been reached in YbRh$_2$Si$_2$ \cite{Geg08.2} and for CeCoIn$_5$, the rise in the c-axis linear resistivity is more than 100 times the residual resistivity of $0.4\,\mu\Omega$cm \cite{Tan07.1s}.

\subsection*{Fermi volume jump}
If the linear-in-temperature resistivity were to be associated with an orbital-selective Mott transition, this should leave traces in Hall effect measurements. In the case of heavy fermion metals, sharp statements can be made: In the fully Kondo-screened Fermi liquid ground state of a Kondo lattice, the local moments contribute to the Fermi volume, which is thus ``large'' \cite{Osh00.1}. Across a Kondo-destruction quantum critical point, the static Kondo screening breaks down and the local moments will thus be expelled from the Fermi sea, leading to a jump of the Fermi volume at $T=0$ \cite{Si01.1,Col01.1,Sen04.1}. The associated jump of the charge carrier concentration is illustrated in FIG.\,\ref{fig3}f.

This behavior was studied in detail in YbRh$_2$Si$_2$. It was found that (linear-response) Hall coefficient vs magnetic field isotherms display a crossover that sharpens with decreasing temperature, extrapolating in the $T=0$ limit to a jump at the material's quantum critical point \cite{Pas04.1,Fri10.2} (FIG.\,\ref{fig3}g). Finite-temperature crossovers were also observed in other heavy fermion compounds, e.g.\ in CeCoIn$_5$ upon Cd and Sn doping \cite{Mak22.1} (FIG.\,\ref{fig3}h), in a study combining several doped cuprates \cite{Bad16.1} (FIG.\,\ref{fig3}i), and hinted at in further materials \cite{Fan22.1,Hua22.1,Wak23.2}. While a final conclusion on whether or not the crossovers in these other materials extrapolate to $T=0$ jumps, perhaps smeared by disorder, is pending, these results are at least suggestive of similar physics at play. On the physics underlying the transition in the cuprates, there is no consensus. Clearly, an orbital-selective Mott transition within the metallic part of the phase diagram requires two types of low-energy electronic degrees of freedom, as discussed further in Sec.\,\ref{Unify}).

In MATBG an anomaly is observed in the Hall effect vs total charge (or gate voltage) isotherms, centered at half-filling and sharpening substantially with decreasing temperature \cite{Cao18.1} (FIG.\,\ref{fig3}j). This filling seems to be associated with some kind of ``Mott'' insulating state. Interestingly, by bringing a screening layer in close vicinity to the twisted bilayers, this insulating state is quenched and a strange metal results \cite{Jao22.1}. Further investigations are needed to evaluate whether these phenomena can be understood in terms of an orbital-selective Mott transition.

\subsection*{Dynamical scaling}
If the orbital-selective Mott transition is a continuous quantum phase transition, quantum critical fluctuations and the associated dynamical scaling should be observed. As it involves a charge localization--delocalization transition, it should be detectable by dynamical scaling in the optical conductivity (FIG.\,\ref{fig3}k, left axis). In heavy fermion compounds, the transition involves the break up of the Kondo singlet, localizing the $f$ electrons. The resulting lattice of magnetic moments may undergo a magnetic ordering transition unless this is prevented by effects such as disorder or frustration. In fact, the dynamical scaling in the spin susceptibility, with a fractional critical exponent (FIG.\,\ref{fig3}k, right axis)---observed in inelastic neutron scattering experiments on quantum critical CeCu$_{5.9}$Au$_{0.1}$ \cite{Sch00.1} (FIG.\,\ref{fig3}l) and similarly in UCu$_{5-x}$Pd$_x$ \cite{Aro95.1}---has played an essential role in defining the phenomenon of Kondo destruction quantum criticality. The same kind of scaling was also found in the heavy fermion semimetal CeRu$_4$Sn$_6$ \cite{Fuh21.1}, which is genuinely quantum critical (FIG.\,\ref{fig3}m). In the strange metal regime of YbRh$_2$Si$_2$, whereas neutron data of sufficient quality to allow for a scaling analysis are not yet available, the optical conductivity exhibits dynamical scaling with a critical exponent of 1, consistent with the linear-in-temperature (dc) resistivity \cite{Pro20.1} (FIG.\,\ref{fig3}n). Closely related is the observation of field-over-temperature ($H/T$) scaling in the strange metal phase of YbAlB$_4$ \cite{Mat11.1} which exhibits a critical charge mode in its low-temperature Moessbauer spectra \cite{Kob23.1}. We note that strong deviations from a Drude-like optical conductivity are a hallmark of strange metals \cite{Pro20.1,Li23.1}. Optical conductivity data on the cuprate La$_{1.76}$Sr$_{0.24}$CuO$_4$ \cite{Mic23.1} show the same scaling behavior if, as also done in \cite{Pro20.1}, a residual optical conductivity is taken into account \cite{Li23.1} (FIG.\,\ref{fig3}o). Associated with the dynamical scaling is a linear-in-$\omega$ and linear-in-$T$ relaxation rate \cite{Pro20.1,Li23.1}. Evidence for this behavior has come from ARPES measurements of the single-particle spectral linewidth in both the cuprates \cite{Val00.1} and the flat band system Fe$_3$Sn$_2$ \cite{Eka22.1x}. The singular charge responses at the localization--delocalization transition also play a key role in the unconventional superconductivity driven by it, with transition temperatures that are high as measured by the effective Fermi temperature \cite{Hu21.2x}.

\subsection*{Loss of quasiparticles}
The linear-in-temperature resistivity, Fermi volume jump, and dynamical energy over temperature scaling discussed above are all at odds with Fermi liquid theory. As such they might hint at the absence of well defined (Fermi-liquid-like) quasiparticles. Shot noise measurements probe the fate of charge carriers in a solid more directly and can thus give a definite answer. Such experiments were recently performed on the heavy fermion compound YbRh$_2$Si$_2$ in its strange metal regime \cite{Che23.1}. The current noise power density across a diffusive wire, fabricated by patterning a thin film grown by molecular beam epitaxy \cite{Pro20.1,Bak22.1}, was found to be strongly suppressed, leading to a Fano factor of less than 0.15 at the lowest temperatures reached in the experiment \cite{Che23.1} (FIG.\,\ref{fig3}q). This is much lower than the Fano factors of $1/3$ in a Fermi gas \cite{Bee92.1} and $\sqrt{3}/4$ with even strong scattering in a Fermi liquid \cite{Wan22.2x}, respectively (FIG.\,\ref{fig3}p). The opposite of quasiparticle transport is the flow of a continuous fluid, which has no fluctuations; thus, on general grounds a lack of quasiparticles reduces the shot noise and the associated Fano factor \cite{Che23.1,Wan22.2x}. A recent calculation of shot noise in a model for fermions randomly coupled to equilibrated critical bosons is consistent with such considerations \cite{Nik23.1}. Similar experiments in other materials across the platforms of FIG.\,\ref{fig1} are of great interest to test whether the lack of quasiparticles is a universal characteristic of strange metals.

\section{Different routes to correlated topology}\label{topology}
The fractional quantum Hall insulator provided an existence proof for a topological state of matter that is driven by strong correlations, at least in gapped settings. More generally, candidate quantum spin liquid states illustrate the power of fractionalization and long-range entanglement in producing correlated topology. Recent rapid progress in frustrated lattice, moir\'{e}, and $f$ and $d$ electron systems suggests a potentially systematized approach to this outstanding challenge, with the narrow bands therein being a common thread.

\subsection*{Gapless phases}
Gapless topological phases in the form of Weyl and Dirac semimetals have been discussed in flat band, $f$ electron, and transition metal systems. For flat band systems, this includes a theoretical proposal for Weyl phases that arise from combining time reversal symmetry breaking and spin--orbit coupling in 3D frustrated lattice systems \cite{Zho19.1}, driving new band crossings that might realize minimal Weyl structures with strongly reduced Weyl velocities compared to those of generic non-interacting systems. Similarly, 3D extensions of twisted van der Waals systems have been proposed to realize inversion breaking Weyl states with flat dispersion through chiral structures that realize nonsymmorphic lattices via repeated misorientation steps \cite{Wu20.1}. In both cases, adding even modest $U$ to the models would result in sizable correlation strength $U/W$. Experimentally, the frustrated hopping-derived narrow $d$ states in the kagome metal \ce{Ni3In} have been shown to support heavy-fermion-like behavior in a manifold of bands containing nodal lines and supporting a topological state with spin--orbit coupling \cite{Ye21.1x}, potentially connecting to earlier theoretical studies of correlations and flat bands in kagome systems (top right of FIG.\,\ref{fig4}) \cite{Vol85.1,Maz14.1,Has20.1}. These observations may have implications for the previously studied transition metal compound \ce{LiV2O4} \cite{Kon97.1}. Long discussed as an unusual example of a transition metal heavy fermion system, from the present viewpoint it seems natural to consider the potential role of the underlying V pyrochlore network in driving a relatively flat electronic band akin to those recently reported in pyrochlore intermetallics \cite{Hua23.3x, Wak23.1} (and of a similar role to Ni in \ce{Ni3In}) which may further be topologically non-trivial \cite{Guo09.1}. A complete theoretical description of the emergence of this behavior from the complex underlying band structure is of significant interest, including to guide the design of further such systems.

For transition metal systems, topological semimetallic behavior has been predicted early on for the pyrochlore iridates, driven by the combination of strong spin--orbit coupling from Ir and proximity to Mott behavior \cite{Wan11.2}. Heavy fermion systems provide an opportunity to realize topological semimetals associated with the low-energy electronic excitations that generically emerge from strong correlations. The notion of Weyl--Kondo semimetals was advanced in concurrent experimental \cite{Dzs17.1} and theoretical \cite{Lai18.1} studies (top left of FIG.\,\ref{fig4}). Experiments on \ce{Ce3Bi4Pd3}, which has nonsymmorphic and noncentrosymmetric space group symmetries \cite{Dzs17.1} but preserves time reversal symmetry \cite{Dzs21.1}, found evidence for a strongly correlated Weyl semimetal \cite{Dzs17.1,Dzs21.1}, with nodes positioned within the narrow gap of an underlying Kondo insulator \cite{Dzs22.1}; signatures of this gap were also emphasized in Refs.~\cite{Kus19.1,Tom20.1}. Theoretical analyses of Kondo lattice models show that strong correlations cooperate with such symmetries to produce Weyl nodes in the Kondo-generated low-energy excitations \cite{Lai18.1}, and insight that has recently led to suggesting a broader class of potential material platforms for $f$-electron-based Weyl--Kondo semimetals \cite{Che22.1}. As a salient characteristic of such topological heavy fermions, both experiment and theory provide evidence that the Weyl nodes are pinned to the immediate vicinity of the Fermi level, illustrating the point that single-particle excitations driven by extreme correlations inherently develop near the Fermi energy. The fact that symmetry constraints apply to the eigenstates of the interacting Green's function provides a foundation for the theoretical analysis in strongly correlated settings \cite{Hu21.1x}.

\subsection*{Gapped phases}
Gapped topological phases are also well represented across the systems reviewed here. Topological insulating ($Z_{2}$) behavior has been proposed in line graph compounds wherein the combination of spin--orbit coupling and the kagome network can give rise to topologically non-trivial gaps \cite{Guo09.1}.  Most gapless flat bands can become topologically gapped once spin--orbit coupling is added. It is then natural to expect (supported by recent theoretical material studies  \cite{Reg22.1}) that many gapped topological materials of this type exist. Real materials typically have complex electronic structures---efforts to understand this behavior and further to design and realize ``ideal'' flat band systems is an important direction for future work. Fragile \cite{Po19.1} and stable \cite{Son19.1} topology have also been discussed in the context of twisted bilayer graphene.

Topological insulating gaps driven by band inversion have also been discussed in transition metal compounds such as (the normal state of) Fe(Se,Te) \cite{Wan15.1} and the transition metal oxide Sr$_2$FeMoO$_6$ \cite{Cha22.1}. A heavy fermion analog of band inversion has been widely discussed in the cubic Kondo insulator SmB$_6$ (middle left of FIG.\,\ref{fig4}), wherein topological insulating behavior is thought to occur via an inversion of light $d$ bands and flat $f$ bands at the three high symmetry X points \cite{Dze16.1}. In such systems the topological gap is brought to the Fermi level via interactions (akin to behavior of gapless heavy fermion systems), overcoming the significant challenge of energetic control of such states in other systems.

Magnetic field-free lattice-based analogs of fractional quantum Hall states---dubbed fractional Chern insulators (middle right of FIG.\,\ref{fig4})---have further demonstrated the power of gapped phases in enabling sharp probes of their underlying topology even in the presence of strong electronic correlation. Decade-old predictions of fractional Chern phases in a variety of lattice models \cite{Neu11.1,She11.1,Reg11.1,Sun11.2,Tan11.1}, once thought to be unrealizable, have seen dramatic recent progress in experiments with twisted bilayer graphene \cite{Xie21.1}, \ce{MoTe2} van der Waals heterostructures \cite{Cai23.1,Zen23.1,Par23.1}, and rhombohedral pentalayer graphene \cite{Lu23.1x} (see Sect.\,\ref{TopSig}). A natural hypothesis is that similar behavior may arise in $f$ electron and transition element compounds, though specific candidate systems have not yet been identified. Progress in this direction may require methods beyond density functional theory, which has driven rapid progress in noninteracting and weakly interacting systems.

\subsection*{Topological superconducting phases}
The study of topological superconductivity is an ongoing effort across the materials platforms. In transition metal and heavy fermion materials, the narrowness of the bands gives rise to an enhanced role of interactions as drivers of strong-coupling pairing gaps, and spin--orbit coupling, an important driver of topological ground states. Candidate heavy fermion topological superconductors include UTe$_2$ \cite{Ran19.1,Jia20.2,Aok22.1} and CeRh$_2$As$_2$ \cite{Khi21.1}. A well-studied candidate among the transition metal superconductors is Fe(Se,Te) \cite{Yin15.1,Zha18.2,Wan18.1,Mac19.1}, potentially representing a correlated realization of the Fu-Kane model for topological superconductivity (bottom left of FIG.\,\ref{fig4}) \cite{Fu08.1}.

Flat band systems, along with their topology, are ideal playgrounds for superconductivity: their large density of states implies large strong-coupling pairing gaps, while their topology gives rise to extended states that can carry the superfluid weight. This has inspired a number of theoretical proposals, including ones for doped magnetic moir\'e heterostructures \cite{Cre23.1} and three-band touching systems \cite{Lin20.1} as potential drivers for topological superconducting states (bottom right of FIG.\,\ref{fig4}). Experimentally, topological superconductivity is for instance pursued in the context of ternary kagome systems \cite{Jia23.1}. The recently discovered transition metal dichalcogenide superconductor 4Hb-TaS$_2$ \cite{Rib20.1}, consisting of alternating layers of insulating 1T-TaS$_2$ and superconducting 1H-TaS$_2$, is also a candidate flat band topological superconductor. Hybridization between the alternating layers gives rise to a superconductor with signatures of broken time reversal symmetry in muon spin rotation measurements \cite{Rib20.1}, which exhibits 1D boundary edge modes in STM spectroscopy \cite{Nay21.1}. Superconductivity---topological or trivial---is also promoted by a nontrivial quantum geometry (Fubini--Study metric) of flat bands arising from quantum interference; this effect can be nonzero even for topologically trivial bands and bounds the superfluid weight from below by preventing full localization \cite{Peo15.1,Her22.1}.

\section{Signatures of correlated topology}\label{TopSig}
\begin{figure}[ht!]
\centering
\includegraphics*[width=0.8\textwidth]{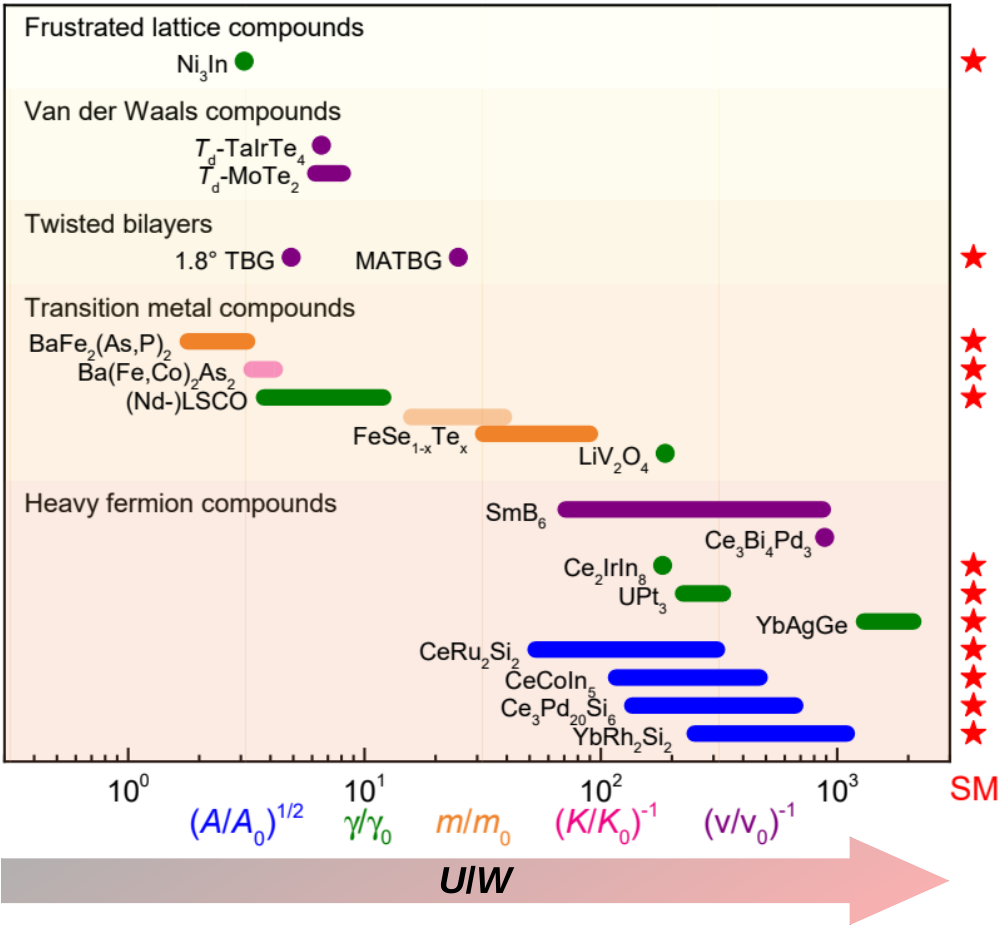}
\caption{\label{fig2} {\bf Experimental measures of correlation strength across the materials platforms.} Values or ranges of the scaled parameters $(A/A_0)^{1/2}$, $\gamma/\gamma_0$, $m/m_0$, and $(K/K_0)^{-1}$ for Schr\"odinger-like quasiparticles and $(v/v_0)^{-1}$ for Dirac-like quasiparticles, extracted as explained in the text from published data of exemplary correlated materials (see below). When ranges are shown, literature data under non-thermal control parameter tuning were available. In this case, strange metal (SM) behavior frequently arises in the vicinity of the largest band renormalizations (see star symbol). The $(A/A_0)^{1/2}$ and/or $\gamma/\gamma_0$ values correspond to $m/m_0$ values, listed in Table 1 of \cite{Tau22.1} for the heavy fermion compounds in different magnetic field ranges, plotted in Figure 3a of \cite{Leg19.1} for the cuprates La$_{2-x}$Sr$_x$CuO$_4$ (LSCO) and La$_{1.6-x}$Nd$_{0.4}$Sr$_x$CuO$_4$ (Nd-LSCO) for p-type doping from 0.24 to 0.4, and from data in \cite{Ura00.1} for LiV$_2$O$_4$ and \cite{Ye21.1x} for Ni$_3$In. $m/m_0$ values of FeSe$_{1-x}$Te$_x$ ($0.56 < x < 0.89$) are extracted from parabolic fits to the $d_{xy}$ band imaged by ARPES; for completeness, we also present $m/m_{\rm DFT}$ data as the shaded bar \cite{Hua22.1}. $m/m_0$ values of BaFe$_2$(As$_{1-x}$P$_x$)$_2$ ($0.4 < x < 0.7$) are determined from dHvA measurements \cite{Shi10.2}. $(K/K_{\rm DFT})^{-1}$ for Ba(Fe$_{1-x}$Co$_x$)$_2$As$_2$ ($0.05 < x < 0.11$) is extracted from the Drude spectral weight \cite{Deg11.1}. $(v/v_0)^{-1}$ is determined from specific heat for Ce$_3$Bi$_4$Pd$_3$ \cite{Dzs17.1}, from STM for SmB$_6$ \cite{Pir20.1}, from SdH oscillations for MATBG \cite{Cao18.2}, 1.4$^{\circ}$ bilayer graphene \cite{Cao16.1}, and $T_d$-MoTe$_2$ \cite{Zho18.1,Kir24.1}, and from ARPES for $T_d$-TaIrTe$_4$ \cite{Hau17.2,Kir24.1}.}
\end{figure}

\begin{figure}[ht!]
\centering
\vspace{-1cm}

\includegraphics*[width=0.95\textwidth]{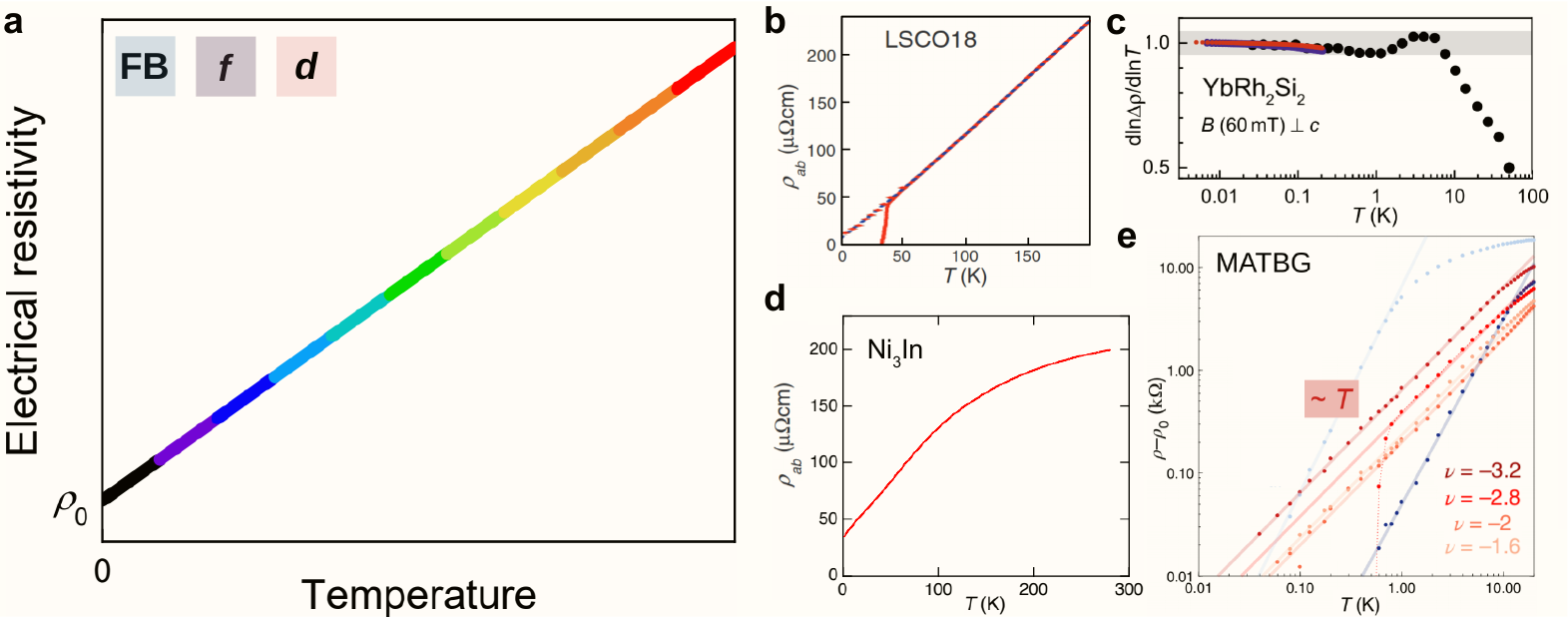}
\vspace{0.1cm}

\includegraphics*[width=0.95\textwidth]{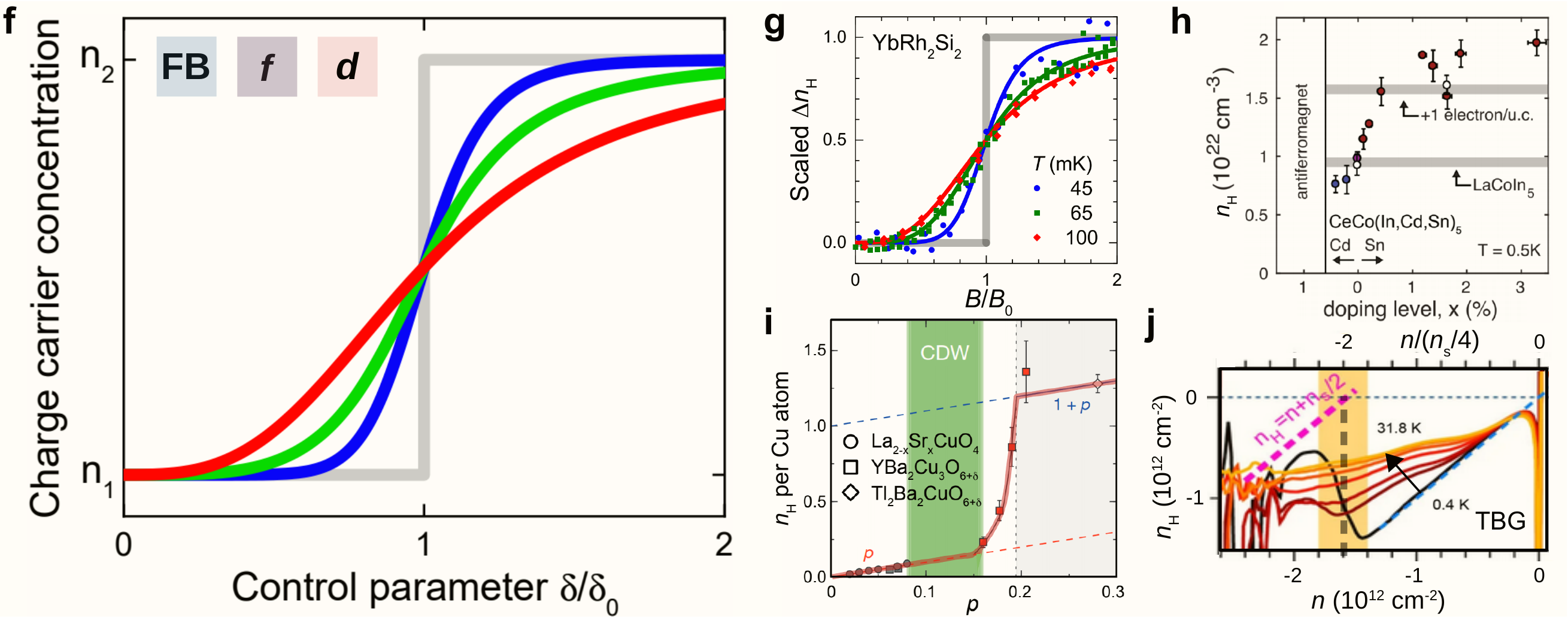}
\vspace{0.1cm}

\includegraphics*[width=0.95\textwidth]{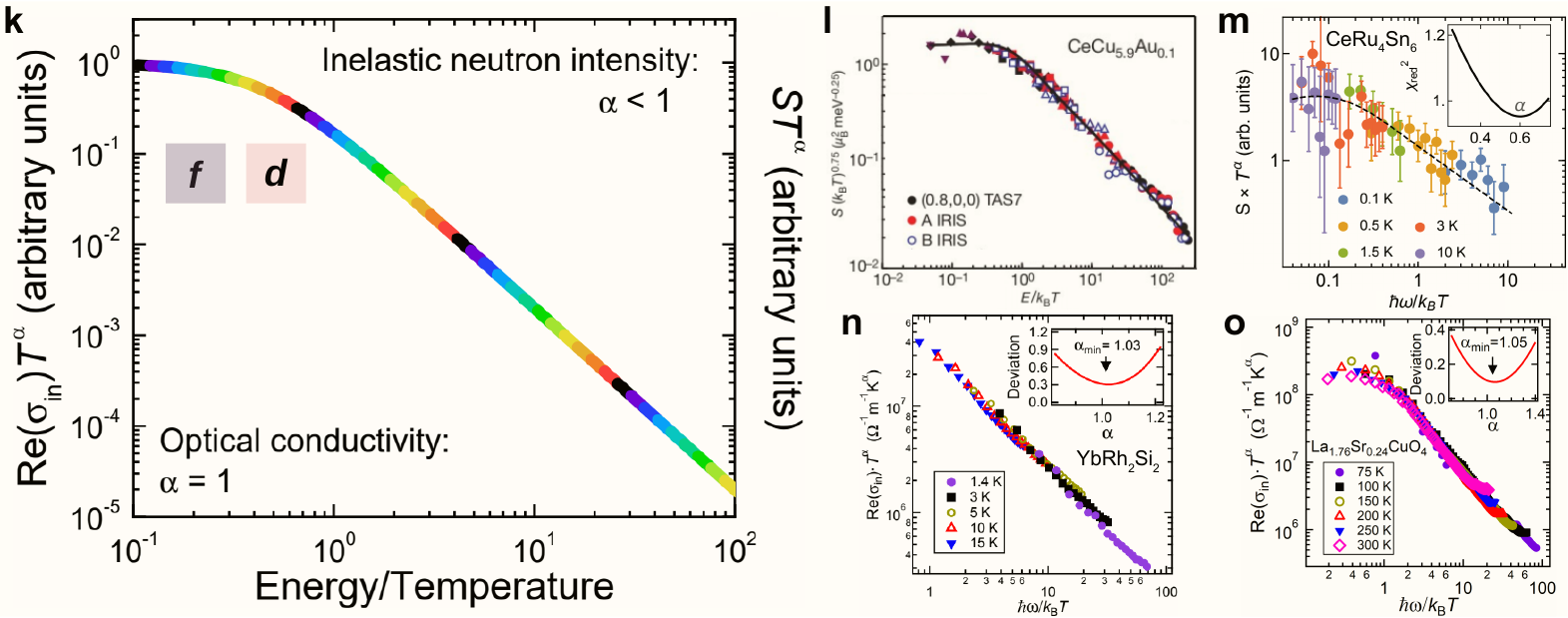}
\vspace{0.1cm}

\includegraphics*[width=0.95\textwidth]{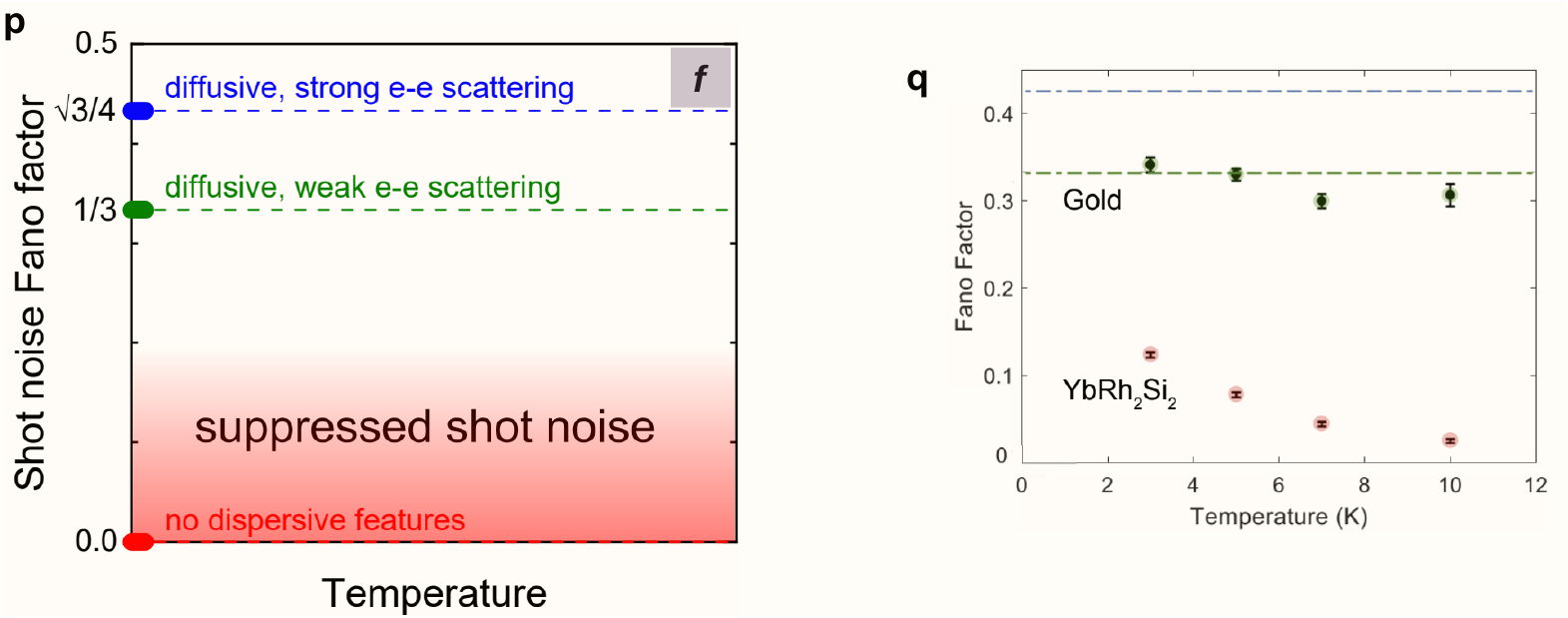}

\caption{\label{fig3} {\bf Strange metal phenomena in the extreme correlation regime.}}
\end{figure}

As seen from FIG.\,\ref{fig2}, topological bands in correlated electron systems are narrower---sometimes by orders of magnitude---than those in non-interacting generic systems. Furthermore, correlations also lead to quasiparticle lifetime broadening. Both effects, together with the lack of precise ab initio predictions for strongly correlated materials, impede the spectroscopic identification of topological signatures of strongly interacting systems. On the other hand, the reduced electronic bandwidth and associated increased electronic density of states offer new possibilities to amplify topological responses. Thermodynamic probes become feasible and transport signatures may be strongly enhanced. A few exemplary signatures of correlated topology are highlighted in FIG.\,\ref{fig5}, grouped into signatures of correlated topological insulators, semimetals, and superconductors.

In the Kondo insulator SmB$_6$, surface states have been identified in a wide range of experiments. They exhibit a robust electrical conductivity \cite{Wol13.1}, despite a highly insulating interior \cite{Eo18.1}; moreover, ARPES measurements have identified Dirac cones on both the 100 \cite{Xu14.1} and 111 surfaces \cite{Oht19.1} with spin textures in accord with expectation \cite{Leg15.1}. Quasiparticle interference patterns in low-temperature STM experiments have imaged extremely flat linearly dispersing bands (FIG.\,\ref{fig5}a); the small Dirac velocities shown in FIG.\,\ref{fig2} were extracted from their slopes. The features are fuzzy, likely due to quasiparticle lifetime broadening. One of the enduring mysteries of this topological flat band system concerns the observation of a large linear specific heat and bulk dHvA oscillations \cite{Tan15.1}. Various viewpoints have arisen: in one the oscillations are interpreted as the damped remnant of underlying $d$ bands in a gapped insulator, arising when the cyclotron energy $\hbar \omega_c$ becomes comparable with the gap (measured to be about 3\,meV) \cite{Kno17.1}; another view \cite{Ert17.1,Sod18.1} is that the Lifschitz--Kosevich temperature dependence of the oscillations at temperatures far smaller than the gap indicates the presence of a gapless Fermi surface of ``neutral'' $d$-like excitations; finally, in Ref.\,\cite{Pir23.1} it is argued that the observed oscillations may derive from non-percolating metallic puddles that form around magnetic impurities.

In gapped twisted bilayer systems, efforts have focused on realizing the fractional quantum anomalous Hall effect (QAHE), the zero-applied-magnetic-field analog of the fractional QHE \cite{Sto99.1} in crystals \cite{Kli20.1}, and thereby evidencing a zero-field fractional Chern insulator (FCI). In MATBG, traces in the Landau fan diagram associated with fractionally quantized Hall conductance disappear below about 5\,T \cite{Xie21.1}, thus coming close but ultimately failing to realize this state. Indirect evidence for a fractional QAHE was provided in twisted bilayers of MoTe$_2$, via the magnetic field dependence of gaps revealed by minima in the photoluminescence intensity and interpreted via the Streda formula \cite{Cai23.1}, and via a combination of local electronic compressibility and magneto-optical measurements \cite{Zen23.1}. Most recently, initial difficulties to create electrical contacts on MoTe$_2$ layers were overcome and the first transport evidence was provided (FIG.\,\ref{fig5}b) \cite{Par23.1} (see also recent work on rhombohedral pentalayer graphene \cite{Lu23.1x}). As these materials are ferromagnets and display hystereses in the above quantized properties, a small magnetic field is still needed to polarize them.

In correlated topological semimetals, narrow bulk bands with reduced Dirac or Weyl quasiparticle velocity are expected and have indeed been evidenced (see $(v/v_0)^{-1}$ in FIG.\,\ref{fig2}). In the bulk (3D) Weyl--Kondo semimetal Ce$_3$Bi$_4$Pd$_3$, the Weyl band is so flat that its specific heat signature, $C \sim T^3$ \cite{Dzs17.1,Lai18.1}, even overshoots the material's phonon contribution \cite{Dzs17.1} (FIG.\,\ref{fig5}c). Furthermore, the nonlinear Hall effect predicted for noncentrosymmetric Weyl semimetals in the absence of time reversal symmetry breaking \cite{Sod15.1} is extremely large \cite{Dzs21.1} (FIG.\,\ref{fig5}d).

To ascertain topological superconductivity experimentally continues to be a formidable challenge. In transition metal compounds, ARPES measurements have detected the presence of topological surface Dirac cones in the normal-state of Fe(Se,Te) \cite{Zha18.2}, which may give rise to a  topological superconductor below $T_{\rm c}$. This is also suggested by STM investigations which observe a clean superconducting gap, together with zero-bias anomalies in the vortex cores that unlike conventional, non-topological Caroli--de Gennes--Matricon bound states (CBSs) \cite{Car64.1}, do not Zeeman-split in a magnetic field \cite{Yin15.1,Wan18.1}.  Indeed, high-energy-resolution STM  measurements \cite{Mac19.1} have resolved a clear separation between the zero-bias anomalies and the CBSs (FIG.\,\ref{fig5}e). In a 45$^{\circ}$ twisted bilayer of the cuprate Bi$_2$Sr$_2$CaCu$_2$O$_{8+x}$, half-integer Shapiro steps indicate the co-tunneling of Cooper pairs, a necessary ingredient for high-temperature topological superconductivity \cite{Zha21.1x}. As to heavy fermion superconductors, tentative evidence for topological superconductivity was found for UTe$_2$, in the form of chiral surface states within the superconducting gap \cite{Jia20.2} (FIG.\,\ref{fig5}f).

\section{Towards a unified understanding}\label{Unify}
Different pathways to strong correlation phenomena, both in the presence and in the absence of electronic topology exist in quantum and engineered materials. Here we discuss the emerging view that the dual itinerant and localized nature of the electrons across the considered materials platforms and their interaction, captured by new forms of effective extended Anderson or Kondo Hamiltonians, can provide a unified understanding of their physics.

\subsection*{Kondo lattices}
The case of heavy fermion compounds, where this dual nature is well understood, provides guidance. Kondo lattice behavior arises most typically in rare earth (or actinide)-based intermetallics, which contain both localized $4f$ (or $5f$) electrons and itinerant $s$, $p$, and $d$-derived electrons (FIG.\,\ref{fig6}a). At high temperatures, the localized and itinerant electrons are largely independent. The localized electrons form local moments, understood in terms of the Hund's rules of atomic physics, while the itinerant electrons form broad bands, well described by single particle theories such as DFT. With decreasing temperature, however, both subsystems start to hybridize. The conduction electrons at first scatter incoherently from the local moments and, with further decreasing temperature, become entangled with the local moments to form new ``heavy fermion'' quasiparticles. The emergent bands are extremely narrow (or ``flat'') and pinned to the Fermi energy \cite{Hew97.1}. More recently it has been realized that, even in the zero temperature limit, these heavy quasiparticles retain their dual nature and are susceptible to breaking apart, a phenomenon referred to as Kondo destruction \cite{Kir20.1,Pas21.1}. It is at the heart of a wealth of exotic phenomena, frequently collectively referred to as ``strange metallicity''.

This physics does not require any particular lattice symmetry nor otherwise generated electronic topology. Yet, it is not incompatible with it, as shown by recently evidenced examples of topological Kondo insulators, semimetals, and superconductors. In fact, the strong spin--orbit coupling of heavy atoms that typically constitute these materials, together with the great flexibility in the structure design of ternary or even quarternary intermetallics, make heavy fermion compounds a versatile platform for new correlation-driven phases. In particular, studies of the recent past have revealed a topological interplay between the localized and itinerant degrees of freedom, demonstrating how crystalline symmetries constrain the Kondo-driven low-energy excitations and create strongly correlated topological states of matter.

\subsection*{Orbital-selective Mott systems}
The optimally hole-doped cuprates (FIG.\,\ref{fig6}b) share strikingly similar properties with quantum critical heavy fermion metals, which suggests a shared description. Especially revealing is the experimental indication for a large-to-small Fermi surface transformation as the hole doping is reduced across the optimal doping \cite{Bad16.1,Fan22.1}. This behavior may very well be common among doped Mott insulators, as a similar Fermi surface transformation has been indicated in the organic charge transfer salts \cite{Wak23.2}. We envision that strong correlations in a doped Mott insulator divide electrons into a coherent and an incoherent part. The latter serves as the analog of the $f$ orbital of a Kondo lattice: when it contributes to the electronic fluid, a large Fermi surface enclosing $1+x$ holes per unit cell ensues; when it does not, a small Fermi surface enclosing $x$ holes per unit cell develops.

Such an orbital-selective Mott transition becomes microscopically more transparent in iron-based superconductors \cite{Joh11.1,Si23.1}, in which multiple $3d$ orbitals lie near the Fermi energy \cite{Yi17.1,Si23.1}. The orbital-selective Mott transition has been most strikingly demonstrated in the case of Fe(Te,Se) \cite{Hua22.1}. In microscopic models for Fe-based superconductors, an important characteristic dictated by crystalline lattice symmetry is that inter-orbital hopping amplitudes are generically nonzero \cite{Yu13.1,Yu17.1}. This introduces an inter-orbital hybridization, in contrast to conventional models of orbital-selective Mott transition in which the orbitals are assumed to be kinetically decoupled (i.e., with Hubbard-like interactions defined in the band basis instead of the orbital basis) \cite{Ani02.1}. The orbital-selective Mott transition in models of Fe-based superconductors, containing both the Hubbard and Hund's couplings, takes place in the presence of the inter-orbital hybridization \cite{Yu13.1,Yu17.1,Yas17.1}. Because of the presence of the hybridization, this type of orbital-selective Mott transition bears considerable analogy to what happens in the Kondo destruction transition of Kondo lattices, where the $f$-$c$ hybridization/Kondo coupling is dynamically suppressed by competing interaction processes \cite{Si01.1,Col01.1,Sen04.1}.

Fe-based superconductors also allow for iso-electronic tuning at commensurate fillings \cite{Dai09.1,Shi14.1}. This presents an opportunity to approach an orbital-selective Mott transition by tuning the correlation strength at a fixed filling. Its one-orbital counterpart would be the bandwidth-controlled (as opposed to filling-tuned) Mott transition \cite{Ima98.1}. A theory of continuous Mott transition has been formulated  \cite{Sen08.1}, with the Kondo destruction quantum critical point of Kondo lattices \cite{Si01.1,Col01.1,Sen04.1} serving as the motivation and analogy.

\subsection*{Moir\'e to Kondo mapping}
The physics of MATBG---the best known example of a moir\'e material---contains many puzzling features. Experimentally, quantum-dot-like behavior observed in STM studies \cite{Xie19.1}, with a large Coulomb energy scale, suggests the presence of localized electrons. Flat bands have long been associated with localized electrons; this, however, is not the case in twisted bilayer graphene where the flat bands exhibit strong topology \cite{Son19.1,Po19.1} at any small twist angle. The observation \cite{Roz21.1,Sai21.1} of a large entropy in the ordered phase that disappears under a magnetic field suggests the presence of loosely coupled local moments. Superconductivity and many transport experiments \cite{Cao18.1,Lu19.2,Sha19.1,Cho20.1x,Lu21.1,And21.1}, on the other hand, suggest the presence of delocalized electrons.

The symmetries (particle--hole and $C_{\rm 2T}$) of the idealized continuum model \cite{Bis11.1} support a large symmetry group ($U(4)$) \cite{Kan18.2,Bul20.1,Ber21.1,Xie20.1}. At zero strain, integer filling correlated insulator states (some with Chern numbers \cite{Sha19.1,Cho20.1x,Lu21.1}) can be thought of as ferromagnets of this group. In strained samples, another related state, a special translational symmetry broken state \cite{Bul20.2} is expected to win. However, an understanding of the spin entropy \cite{Roz21.1,Sai21.1} and of the Hubbard bands seen in STM experiments \cite{Xie19.1,Cho19.1}, even above the ordering temperature, is impossible within the momentum space description. The topology of the flat bands preempts a lattice, localized description of MATBG \cite{Son19.1,Kan18.2,Kan19.2,Kos18.1,Po19.1,Vaf21.1,Zou18.1}. Projecting the interaction into the momentum space Bloch states of the flat bands \cite{Zha21.1,Bul20.1,Cea20.1,Ber21.1,Xie20.1} results in the loss of any real-space description, and therefore of Hubbard bands.

To obtain a local description while maintaining the symmetries of the idealized model, and to resolve the localized vs delocalized electron dichotomy, one needs to adopt a fundamentally different point of view. MATBG can be mapped \cite{Son22.1} to a new kind of heavy fermion system (FIG.\,\ref{fig6}c). Most ($>92\%$) of its flat bands consist of local orbitals (called ``$f$'', although of $p_x \pm i p_y$ symmetry) centered at the AA-stacking regions \cite{Son22.1}. Their kinetic energy is vanishingly small ($\sim 0.1$\,meV) \cite{Son22.1}, and their on-site Coulomb repulsion is very strong ($\sim 60$\,meV). The delocalized part of the flat bands are plane-wave-like topological conduction bands (``$c$'') which resemble those of AB stacked Bernal graphene; they are equivalent to a double-Dirac node, have no real-space local description, and carry the topology of the problem. The coupling between $f$ and $c$ electrons is computed to be $\sim 20$\,meV, and gives rise to the single-particle gaps of the flat to remote bands of the Bistritzer--MacDonald (BM) model \cite{Bis11.1}. After this hybridization, remarkable agreement with the flat and remote bands of the BM model is obtained \cite{Son21.1,Shi22.1}. The interacting Hamiltonian has a regular ``Anderson'' lattice of $f$ electrons subject to Hubbard $U$, but in addition exhibits a new ferromagnetic $f-c$ spin coupling; this, as well as the topological nature of the $c$ electrons, differentiates this topological heavy fermion model from the regular periodic Anderson model.

Depending on the filling of the flat MATBG bands, which in experiments can be controlled by gates, different correlated states appear. Due to the strong on-site repulsion, the $f$ electrons tend to order at integer fillings, in states that can be found by integrating out the $c$ electrons to obtain $f$ spin--spin models. The charge excitations/fluctuations can be obtained by adding the $c$ electrons and are crucial to the physics of the system away from integer fillings. The coupling between $c$ and $f$ electrons will select the ground state from the (1, 2, or higher-fold) degenerate multiplet formed by $f$ electrons at integer filling. The Chern numbers of the correlated insulator ground states are pre-determined by the coupling to angular momenta of the multiplets formed by $f$ electrons in the many-body state. The STM studies \cite{Xie19.1,Cho19.1} reporting Hubbard bands at higher temperatures also report low-temperature peaks pinned to the Fermi level at fillings away from charge neutrality. These have been theoretically interpreted as Kondo peaks \cite{Hu23.2,Zho23.1x,Hua23.1x,Cho23.1}, although more work is necessary to fully pin down their origin.

As the $f$ electrons are extremely localized, the quantum-dot-like behavior and Hubbard bands observed in STM studies \cite{Xie19.1} are attributed to them, even at high temperatures  \cite{Hu23.1,Shi22.1,Hu23.2,Zho23.1x,Lau23.1x,Hua23.1x,Dat23.1,Cho23.1}. The topological $c$ electrons with unbounded kinetic energies are responsible for the delocalization, Dirac-like behavior \cite{Zon20.1}, and the eventual superconductivity. This resolved the local/itinerant dichotomy and promotes MATBG into a class of novel ``heavy fermion'' systems. 

\subsection*{Frustrated lattice to Kondo mapping}
Electronic materials with frustrated lattices, such as kagome metals, have in recent years been extensively studied for the formation of unusual charge density wave and other electronic orders. When flat bands are active for the low-energy physics, the kagome \cite{Ye21.1x,Eka22.1x} and pyrochlore \cite{Hua23.2x} metals are also emerging as new settings to realize strange metal behavior. Physical properties in these systems are indicative of the presence of local moment fluctuations in a metallic setting. For such frustrated lattices, there is a long-standing notion that a flat band is visualized in terms of compact localized states \cite{Ley18.1,Ber08.1}. This notion suggests a description of a flat band in terms of very localized molecular orbitals. However, the compact localized states are not suitable for this purpose because they do not form a complete orthonormal basis.

The construction of the proper molecular orbitals has been demonstrated in the kagome lattice \cite{Che23.1x} and a simpler variant thereof, the clover lattice \cite{Che22.3x,Hu22.3x}, recognizing that the involved bands are topologically nontrivial. The combination of topological flat and wide bands near the Fermi energy allows for the construction of exponentially localized and symmetry-preserving Wannier orbitals. This is illustrated in a particularly transparent way in a Hubbard model defined on the clover lattice, where a topological flat band develops. The flat band is primarily captured by highly localized (and symmetry-preserving) Wannier orbitals that form a triangular lattice (FIG.\,\ref{fig6}d). These correspond to tight molecular orbitals, i.e. the ones with the shortest decaying length scale. At the same time, the wide bands are mainly represented in terms of the Wannier orbitals with longer decaying length scales, which correspond to more extended molecular orbitals. In contrast to the compact localized states, these molecular orbitals form a complete orthonormal basis.

The tight and more extended molecular orbitals provide the basis in which an effective Anderson/Kondo lattice description emerges. This description captures the selective correlations of the molecular orbitals, when the Coulomb interactions are larger than the width of the flat band and smaller than the width of the wide bands. A continuous orbital-selective Mott transition is shown \cite{Che23.1x,Hu22.3x}, describing metallic quantum criticality where quasiparticles are expected to be lost. This provides a means of understanding the aforementioned strange metallicity in the kagome metals \cite{Ye21.1x,Eka22.1x}. Because the tight molecular orbitals form a frustrated lattice, and by analogy with the global phase diagram of the Kondo lattice systems \cite{Pas21.1}, a quantum critical phase can also develop \cite{Hua23.2x}. Finally, it has been proposed that this represents a route to realize Weyl--Kondo semimetals in $d$-electron-based materials \cite{Che22.3x}.

\section*{Perspectives}
The emerging view is that the paradigm of Kondo physics can serve as a powerful organizing principle for a large range of materials platforms (FIG.\,\ref{fig1}) with correlated electrons  (FIG.\,\ref{fig2}): originally associated with (atomically) localized states coupled to conduction bands, which may or may not be topological, but now extended to flat topological bands introducing a new dichotomy of localized and delocalized electrons living together in the same band (FIG.\,\ref{fig6}). Kondo physics may also be the key to understanding the strange metallicity (FIG.\,\ref{fig3}) seen across the platforms. In heavy fermion compounds, it is attributed to a Kondo destruction transition, also referred to as $f$ orbital selective Mott transition. In the cuprates and iron pnictides, there are indications for such a transition in the $d$ orbital manifold, and in MATBG it may well be occurring within the localized, effective $p_x\pm i p_y$ orbitals. Finally, in $d$ electron systems on frustrated lattices, it can develop via a selective Mott transition of the molecular orbitals when the Coulomb interaction lies in between the widths of the flat and the wide band.

Synergies between the formerly largely disconnected communities on the flat band and strongly correlated electron systems bear great potential to break new ground. Already at the level of terminology, there is scope for unification. As highlighted in FIG.\,\ref{fig2}, ``flat'' bands are not unique to flat band systems but occur in all these platforms, with heavy fermion compounds reaching an extreme band ``flatness'' or correlation strength. On the other hand, while the inherent topological nature of flat bands has been broadly recognized, the field of strongly correlated electron systems has traditionally focused on materials with topologically trivial bands and is only recently identifying topological strongly correlated (``flat'') bands. What do topological phases across the platforms have in common? Can insights from one materials class inspire progress in another? For instance, can the analog of a fractional Chern insulator be realized in Kondo insulators? Can the existence of topological superconductors be firmly established? What is the nature of (even conventional) superconductivity arising in a topological flat band?

An equally exciting open question is what role inherent quantum fluctuations play in stabilizing entirely new topological phases. Whereas unconventional superconductivity is well established to emerge out of a strange metal normal state, there are only first hints that the same might be true for certain Weyl--Kondo phases \cite{Fuh21.1,Hu21.1x}.

There is also high potential for new techniques to boost progress. On the theory side, the understanding of flat band systems as localized electrons coupled to topological semimetals naturally leads to new types of Anderson/Kondo models with physics yet to be discovered. These fronts represent a well-timed confluence with recent developments on topological heavy fermions in which space group symmetries constrain interacting electron excitations. The field may well have reached a juncture when many exciting possibilities can be realized. These include new topological phases driven by a variety of means for strong correlations, an expectation that is lifted by a Green's function way of treating symmetry constraints, and new correlated phases and quantum criticality enriched by topology. The opportunities for cross-talk across the model and materials platforms are enormous. Such prospects for new physics call for---and may well motivate---the development of new methods that can study the interplay of electron correlations and topology non-perturbatively.

Experimentally, signatures beyond those used so far (FIG.\,\ref{fig5}) may help to firmly pin down new correlation-driven topological phases, including topological superconductivity. Quantum spin liquids, a strongly correlated topological phase that we have not explicitly discussed here, are the prototypical systems where ``massive'' many-body entanglement in the ground states is expected \cite{Sav17.1}. Interestingly, a maximally entangled state between a Kondo impurity and the conduction band is expected at an interacting (Kondo destruction-type) quantum critical point \cite{Wag18.1}, a situation that also leads to strange metallicity. Maximal or long-range entanglement may thus be a property that topological and quantum critical systems with fractionalized excitations share. Experimental tools to detect it, including the quantum Fisher information \cite{Hau16.1} but also going beyond it, are in high demand.

\section*{Author contributions}
J.G.C.\ and S.P.\ researched data for the article. All authors contributed to the discussion of the content and the writing and editing of the article before submission.

\section*{Acknowledgements}
We thank J. Analytis, D. Calugaru, J. Cano, L. Chen, R. Comin, L. Crippa, P. Dai, X. Dai, L. Deng, D. Efetov, S. Fang, S. Grefe, P. Guinea, T. Hazra, J. Hertzog Arbeitman, J. Hoffman. H. Hu, J. Huang, K. Ingersent, O. Jansen, P. Jarillo-Herrero, A. Kandala, S. Kirchner, D. Kirschbaum, L. Lau, G. Le Roy, X. Li, G. Lonzarich, M. Lu\v{z}nik, V. Madhavan, M. Mahankali, D. Natelson, J. P. Paglione, A. Panigrahi, J. Pixley, G. Rai, N. Regnault, G. Sangiovanni, S. Sebastian, T. Senthil, C. Setty, Z. Song, F. Steglich, S. Sur, M. Taupin, A. Tsvelik, O. Vafek, R. Valenti, M. Vergniory, Y. Wang, T. Wehling, F. Xie, B. J. Yang, L. Ye, M. Yi, J. Yu, R. Yu, and Z. Zhuang for collaborations and discussions, which were in part conducted at the Kavli Institute for Theoretical Physics at UC Santa Barbara, where support was provided by the US National Science Foundation (NSF) under grant No. NSF PHY-1748958, and at the Aspen Center for Physics, which is supported by NSF grant No. PHY-2210452. J.G.C. was supported by the Gordon and Betty Moore Foundation EPiQS Initiative (Grant No. GBMF9070). B.A.B.'s work was primarily supported by the DOE Grant No. DE-SC0016239 and the Betty Moore Foundation's EPiQS Initiative (Grant No. GBMF11070). Q.S. is primarily supported by the U.S. DOE, BES, under Award No. DE-SC0018197, by the  Robert A. Welch Foundation Grant No. C-1411 and by the Vannevar Bush Faculty Fellowship ONR-VB N00014-23-1-2870. P.C. was supported by NSF grant DMR-1830707. S.P. acknowledges funding from the Austrian Science Fund (I5868-FOR 5249 QUAST, F86-SFB Q-M\&S), the European Union's Horizon 2020 Research and Innovation Programme (824109, EMP), and the European Research Council (ERC Advanced Grant 101055088, CorMeTop).


\clearpage
\newpage

\newpage

\begin{figure*}[ht!]
\noindent cont.\ Figure 3: {\bf Strange metal phenomena in the extreme correlation regime.} {\bf a} Sketch of linear-in-temperature electrical resistivity, as encountered in {\bf b} the cuprate La$_{2-x}$Sr$_x$CuO$_4$ near optimum doping \cite{Coo09.1}, {\bf c} the heavy fermion compound YbRh$_2$Si$_2$ at its quantum critical field \cite{Ngu21.1}, {\bf d} the kagome metal Ni$_3$In \cite{Ye23x}, and {\bf e} MATBG with a closely spaced screening layer at fillings between $\nu = -3.2$ and $-1.6$ \cite{Jao22.1}. {\bf f} Sketch of Fermi volume jump at $T=0$, and its broadening at finite temperatures, as observed in {\bf g} YbRh$_2$Si$_2$ \cite{Pas04.1,Fri10.2}. Rapid crossovers at finite temperature are also seen in {\bf h} CeCoIn$_5$ upon Cd and Sn doping \cite{Mak22.1} and in {\bf i} several cuprates as function of doping \cite{Bad16.1}. {\bf j} A Fermi surface reconstruction also happens in MATBG as function of gating \cite{Cao18.1}. {\bf k} Sketch of dynamical scaling in the optical conductivity (left axis) and inelastic neutron scattering data (right axis), with a critical exponent of 1 and a fractional value $\alpha < 1$, respectively, as observed in {\bf l} the quantum critical heavy fermion metal CeCu$_{5.9}$Au$_{0.1}$ \cite{Sch00.1}, {\bf m} the heavy fermion semimetal CeRu$_4$Sn$_6$ \cite{Fuh21.1}, {\bf n} YbRh$_2$Si$_2$ \cite{Pro20.1}, and {\bf o} the cuprate La$_{1.76}$Sr$_{0.24}$CuO$_4$ \cite{Mic23.1,Li23.1}. {\bf p} Sketch of shot noise suppression, as observed in {\bf q} YbRh$_2$Si$_2$ \cite{Che23.1}. The labels FL (flat band systems), $f$ ($f$-based compounds), and $d$ (transition metal compounds) in the leftmost panels indicate in which materials platform the respective phenomenon has so far been demonstrated.
\vspace{18cm}
\end{figure*}

\clearpage
\newpage

\begin{figure}[ht!]
\centering
\includegraphics*[width=1.0\textwidth]{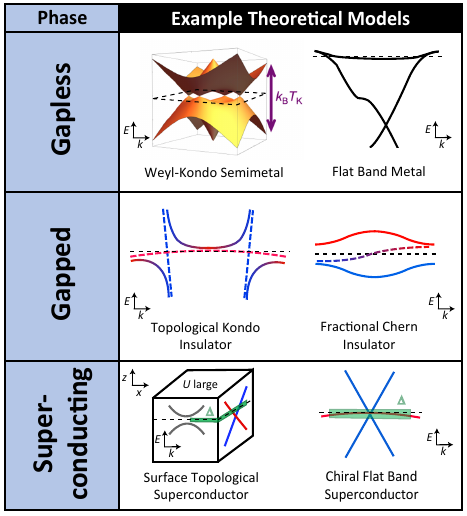}
\caption{\label{fig4} {\bf Flavors of topology in correlated systems.} Exemplary electronic bandstructure cartoons of theoretically predicted topological phases across the materials platforms. For gapless phases, this includes Weyl--Kondo semimetals \cite{Lai18.1}, with nodes pinned to the Fermi energy (dashed plane) and a band flatness given by the Kondo energy scale $k_{\rm B}T_{\rm K}$ (left), and topological flat bands in kagome lattice systems \cite{Vol85.1,Maz14.1,Has20.1} when positioned at the Fermi level (dashed horizontal line, right). Examples of gapped phases are topological Kondo insulators \cite{Dze10.1}, with topological surface states (colored dashed curves) within the hybridization gap (left), and fractional Chern insulators \cite{Neu11.1,She11.1,Reg11.1,Sun11.2,Tan11.1} (right). Correlated topological superconductivity might, for instance, arise on the surface of a topological insulator proximitized with a superconductor with gap $\Delta$ (green shading) \cite{Fu08.1} in systems with large Coulomb interaction $U$ (left), or as chiral bulk superconductivity from three band crossings \cite{Lin20.1} (right).}
\end{figure}

\begin{figure}[ht!]
\centering
\includegraphics*[width=0.95\textwidth]{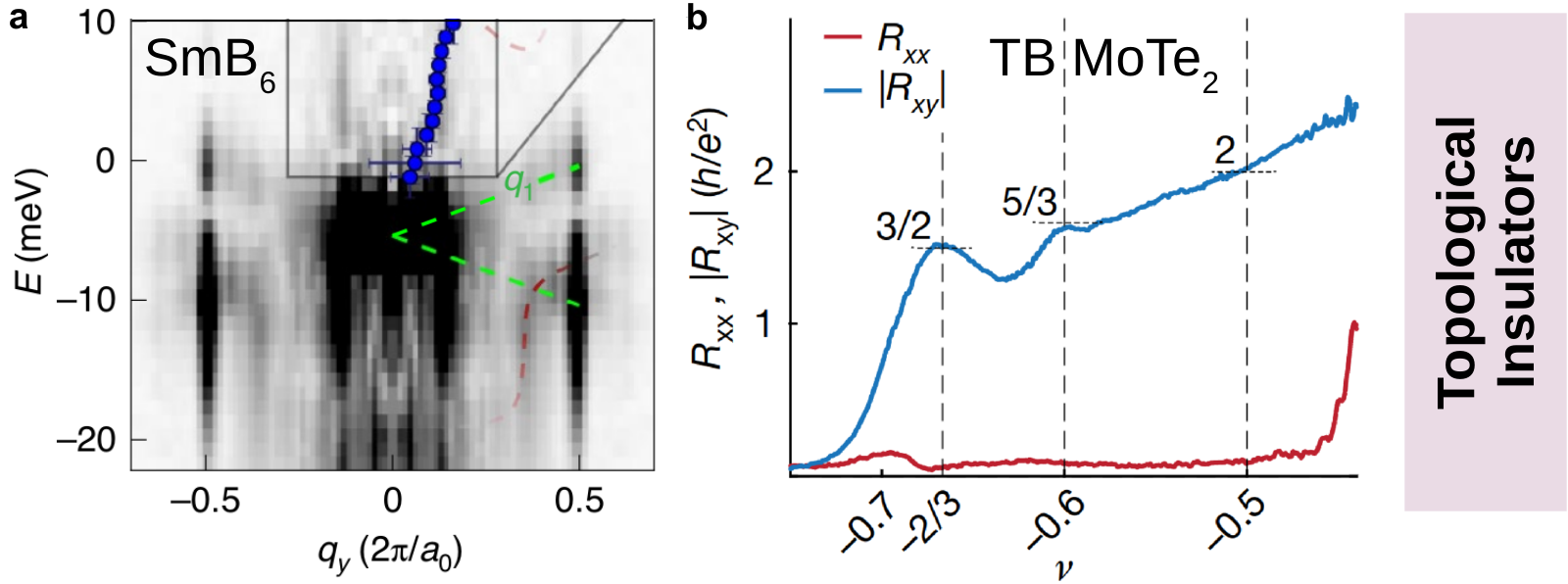}
\includegraphics*[width=0.95\textwidth]{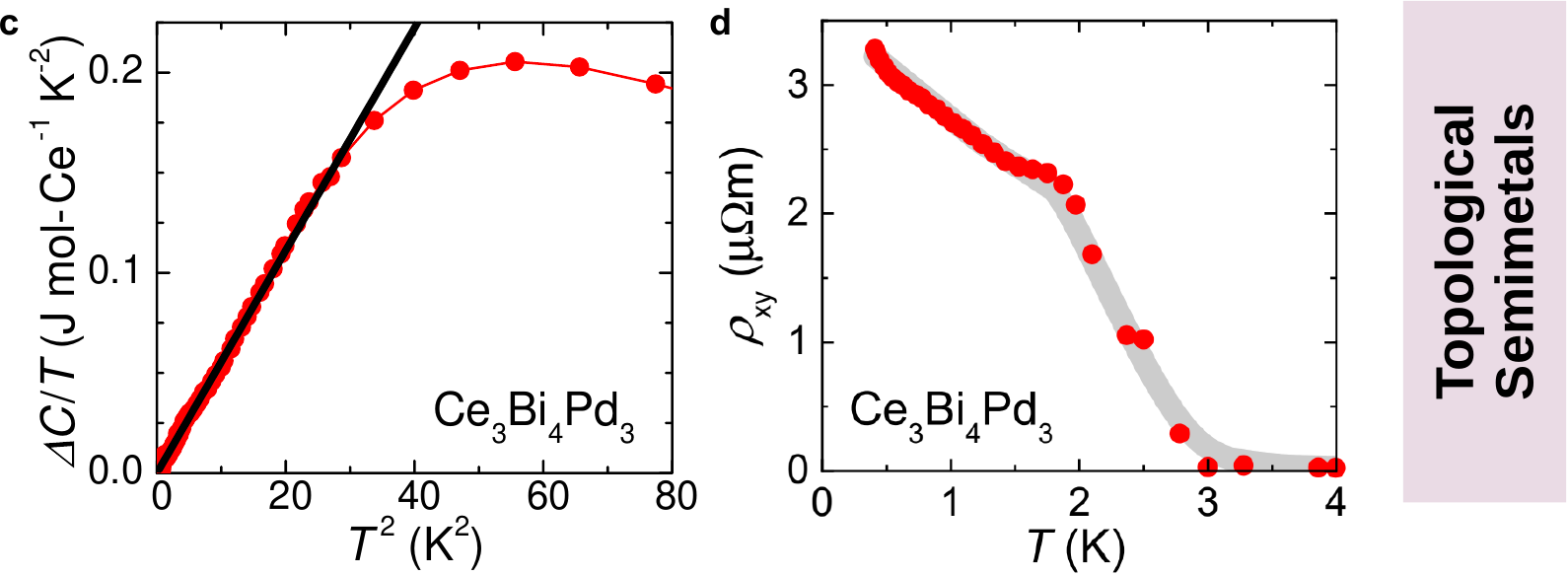}
\includegraphics*[width=0.95\textwidth]{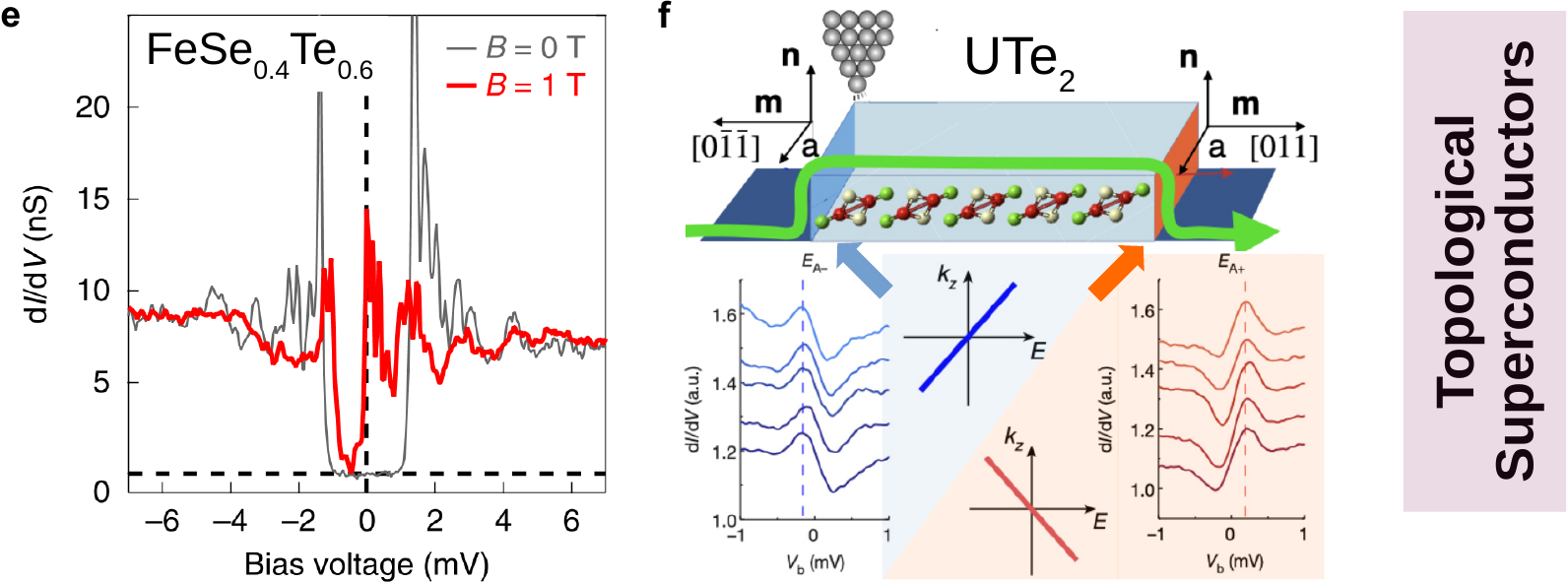}
\caption{\label{fig5} {\bf Signatures of correlated topology---insulators, semimetals, superconductors.} {\bf a} Flat Dirac-like surface states (green dashed lines) on the Kondo insulator SmB$_6$ as evidenced by quasiparticle interference patterns in a low-temperature STM study \cite{Pir20.1}. {\bf b} The longitudinal resistance $R_{xx}$ and absolute value of the Hall resistance $|R_{xy}|$ of twisted bilayer (TB) MoTe$_2$, measured at $\pm 50$\,mT and 100\,mK versus the filling factor $v$, provide transport evidence for a fractionally quantized anomalous Hall effect \cite{Par23.1}. {\bf c} The linear-in-$T^2$ specific heat coefficient $\Delta C/T$ observed in the Weyl--Kondo semimetal Ce$_3$Bi$_4$Pd$_3$ \cite{Dzs17.1} evidences a linear electronic dispersion at the Fermi energy, with a tiny slope (Weyl velocity) \cite{Dzs17.1,Lai18.1}. {\bf d} Giant spontaneous nonlinear Hall effect in Ce$_3$Bi$_4$Pd$_3$, evidencing the fictitious magnetic field monopoles from Weyl nodes in the immediate vicinity to the Fermi energy \cite{Dzs21.1}. {\bf e} Vortex bound states in Fe(Se,Te) as observed by STM, as found for some but not all vortices \cite{Mac19.1}. {\bf f} STM study of UTe$_2$, where the asymmetric line shape of the $dI/dV$ spectra taken at various positions on two sides of a terrace was attributed to chiral surface states within the superconducting gap \cite{Jia20.2}.}
\end{figure}

\begin{figure}[ht!]
\centering
\includegraphics*[width=0.95\textwidth]{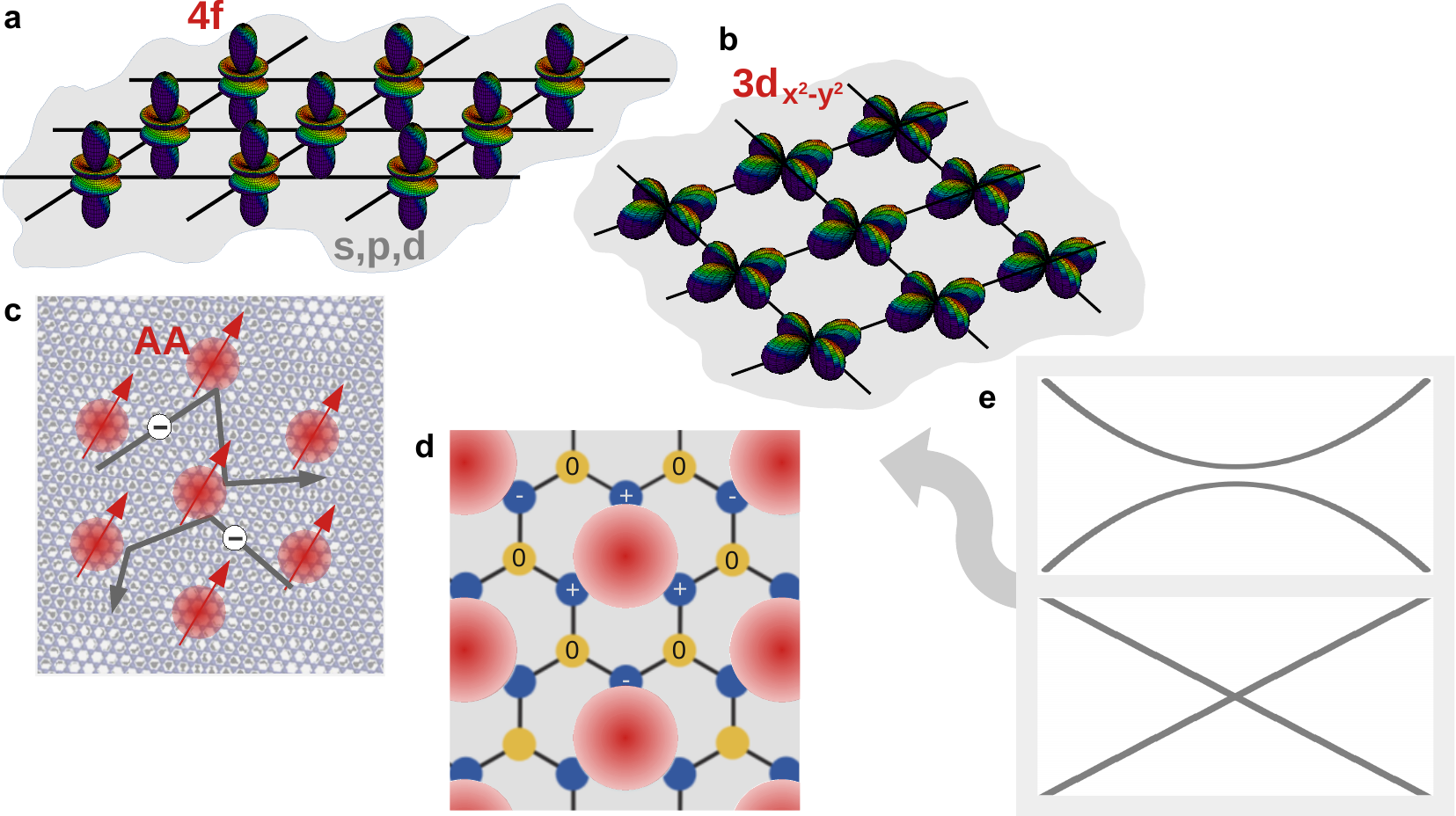}
\caption{\label{fig6} {\bf Heavy fermion physics as a unified view?} {\bf a} The Kondo lattice as the standard model for heavy fermion compounds, with a lattice of localized $4f$ electrons interacting with $s$, $p$, and $d$-derived itinerant electrons. {\bf b} $d$ electron analog of the Kondo lattice, considered for materials (pnictides, cuprates, etc.) with orbital-selective Mott transitions. {\bf c} Heavy fermion mapping of a moir\'e superlattice of MATBG, where the local moments and itinerant electrons are formed by the maximally localized Wannier functions at the AA-stacking regions and topological conduction bands, respectively \cite{Son22.1}. {\bf d} Heavy fermion mapping of the clover line graph lattice, where the Wannier orbitals near the geometric centers of the unit cells represent the local moments \cite{Che22.3x,Hu22.3x}. {\bf e} The itinerant electrons that hybridize with the localized states can be Schr\"odinger or Dirac-like in all cases.}
\end{figure}

\end{document}